\documentclass[letterpaper,english,aps,prb,reprint,superscriptaddress]{revtex4-1}
\usepackage{lmodern}

\usepackage[T1]{fontenc}
\usepackage[latin9]{inputenc}
\usepackage{units}
\usepackage{amsmath}
\usepackage{amssymb}
\usepackage{graphicx}
\usepackage{esint}

\makeatletter

 \@ifundefined{textcolor}{}
 {%
   \definecolor{BLACK}{gray}{0}
   \definecolor{WHITE}{gray}{1}
   \definecolor{RED}{rgb}{1,0,0}
   \definecolor{GREEN}{rgb}{0,1,0}
   \definecolor{BLUE}{rgb}{0,0,1}
   \definecolor{CYAN}{cmyk}{1,0,0,0}
   \definecolor{MAGENTA}{cmyk}{0,1,0,0}
   \definecolor{YELLOW}{cmyk}{0,0,1,0}
 }

\makeatother

\usepackage{babel}
\begin{document}

\title{Temperature Modulation of the Transmission Barrier in Quantum Point
Contacts}

\author{Alfredo X. Sánchez}

\affiliation{Department of Physics and Beckman Institute, University of Illinois
at Urbana-Champaign, Urbana, Illinois 61801, USA}

\author{Jean-Pierre Leburton}

\email{jleburto@illinois.edu}

\affiliation{Department of Physics and Beckman Institute, University of Illinois
at Urbana-Champaign, Urbana, Illinois 61801, USA}

\affiliation{Department of Electrical and Computer Engineering, University of
Illinois at Urbana-Champaign, Urbana, Illinois 61801, USA}
\begin{abstract}
We investigate near-equilibrium ballistic transport through a quantum
point contact (QPC) along a GaAs/AlGaAs heterojunction with a transfer
matrix technique, as a function of temperature and the shape of the
potential barrier in the QPC. Our analysis is based on a three-dimensional
(3D) quantum-mechanical variational model within the Hartree-Fock
approximation that takes into account the vertical depletion potential
from ionized acceptors in GaAs and the gate-induced transverse confinement
potential that reduce to an effective slowly-varying one-dimensional
(1D) potential along the narrow constriction. The calculated zero-temperature
transmission exhibits a shoulder ranging from 0.3 to 0.6 depending
on the length of the QPC and the profile of the barrier potential.
The effect is a consequence of the compressibility peak in the 1D
electron gas and is enhanced for anti-ferromagnetic interaction among
electrons in the QPC, but is smeared out once temperature is increased
by a few tenths of a kelvin.
\end{abstract}
\maketitle

\section{Introduction}

\begin{figure}
\includegraphics[width=2.8in]{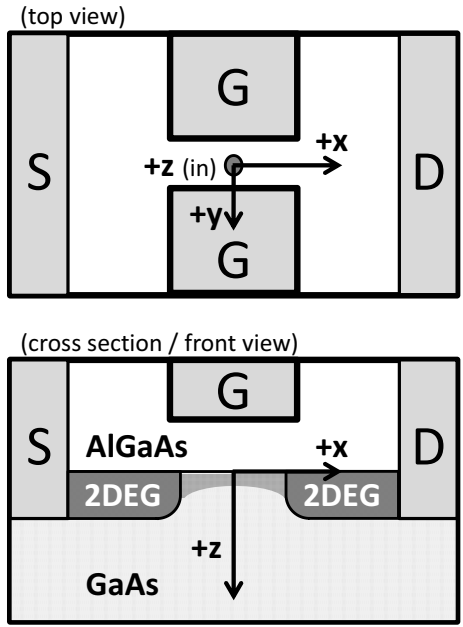}

\caption{\label{fig:1-scheme}Schematic representation of a quantum point contact.
S, D and G stand for source, drain and gate, respectively.}
\end{figure}

A quantum point contact (QPC) is a narrow constriction in a two-dimensional
electron gas (2DEG). QPCs are commonly realized in GaAs/AlGaAs heterostructures
by applying a negative voltage to a split metallic gate placed on
top of the device. This leads to the depletion of the electron gas
directly underneath the gate, leaving only a narrow channel between
two 2DEG reservoirs (acting as source and drain), with a width that
can be modulated by the gate bias (see Figure \ref{fig:1-scheme},
top left). Experiments have shown that the QPC conductance is quantized
in units of\citealp{Wharam1988,Wees1988,Thomas1996} $G_{0}\equiv2e^{2}/h$.
This phenomenon is a direct consequence of quantum-mechanical 1D transmission
through a saddle potential, for which the product of the carrier velocity
and 1D density of states is energy independent. As a result, discrete
steps appear whenever a new 1D channel (1D energy subband) becomes
populated for conduction\citealp{Glazman1988,Buttiker1990}. In addition
to the quantized conductance, an anomalous conductance plateau has
been observed around $0.7G_{0}$\citealp{Thomas1996,Nuttinck2000}.
Of particular interest is the fact that this plateau, dubbed the 0.7
structure or anomaly, becomes stronger as temperature increases above
$\unit[0]{K}$, reaching its maximum strength at $T\sim\unit[1-2]{K}$
and vanishing by $T\sim\unit[10]{K}$ due to thermal smearing\citealp{Thomas1996,Thomas1998}.

There is substantial agreement that the 0.7 structure is a consequence
of many-body effects, but the precise cause of this phenomenon is
still the subject of significant debate\citealp{Berggren2010,Micolich2011}.
One possible explanation involves a Kondo-like effect due to the formation
of a quasi-bound state in the QPC\citealp{Cronenwett2002,Meir2002}.
Alternatively, it has been proposed that a static spin polarization
is present along the constriction\citealp{Thomas1996,Thomas1998,Rokhinson2006}.
However, theoretical calculations based on spin density-functional
theory (DFT) have been inconclusive. While several works favor the
formation of a bound state\citealp{Meir2002,Hirose2003,Rejec2006,Ihnatsenka2007,Meir2008},
others argue against it\citealp{Starikov2003,Berggren2008} and support
the presence of a static spin polarization\citealp{Starikov2003,Havu2004,Berggren2008}.
To add to the controversy, it has been suggested that Kondo correlations
may coexist with static spin polarization due to the presence of localized,
ferromagnetically-coupled magnetic impurity states at the QPC\citealp{Song2011}.

In this work, we re-examine the electron transmission through a QPC
along a GaAs/AlGaAs heterojunction by using a three-dimensional (3D)
self-consistent model that takes into account the lateral confinement
and the potential barrier induced by the gates as well as the band
bending near the GaAs/AlGaAs interface. The system is not treated
as strictly two-dimensional, which allows us to consider the variation
of the vertical confinement on the 2D constriction and the resulting
changes in the conductance. A key issue in our approach is the assumption
that the confinement along the 1D channel is slowly-varying (adiabatic).
For this reason, we first solve the 3D many-body effects of a 1D channel
in the Hartree-Fock approximation to show that tunneling of electrons
through the QPC can be reduced to an effective 1D potential. In our
model, variational wavefunction parameters and the effective potential
are obtained self-consistently as a function of the gate voltage,
confinement strength and temperature. We then use the transfer matrix
technique\citealp{Jonsson1990} to obtain the transmission coefficient
and the conductance through the QPC barrier. We show that, at $T=0$,
the conductance through the QPC exhibits a kink or feature occurring
between $0.3$ and $0.6G_{0}$, depending on the profile of the potential
barrier. The kink is caused by a pinning of the effective potential
when the electron gas in the region of the QPC is depleted, and disappears
as temperature increases.

\section{QPC Structure Model}

A quantum point contact is usually achieved along a GaAs/AlGaAs heterojunction,
which consists of a layer of modulation-doped or delta-doped AlGaAs
on top of a GaAs substrate. The heterojunction interface is about
200-300 nm below the surface\citealp{Thomas1995,Cronenwett2001}.
A cross-section of the structure is shown in Figure \ref{fig:1-scheme}
with the $z$-axis perpendicular to the hetero-interface, oriented
so that $z>0$ measures the depth in the GaAs substrate. Because of
the doping, charge carriers are confined into a 2DEG underneath the
GaAs/AlGaAs heterojunction interface (in the $x-y$ plane) by the
depletion potential $U_{A}$ resulting from the GaAs ionized acceptors.
Meanwhile, the tall hetero-barrier prevents electrons from tunneling
into AlGaAs ($z<0$)\citealp{Ando1982}. Lateral confinement is achieved
along the $y$-direction by applying a negative potential on the split-gate
electrodes (Fig. \ref{fig:1-scheme}). Electron energies are restricted
to discrete subbands which, at zero temperature, will be occupied
only if their energies are below the Fermi energy.

\begin{figure*}
\includegraphics[width=4in]{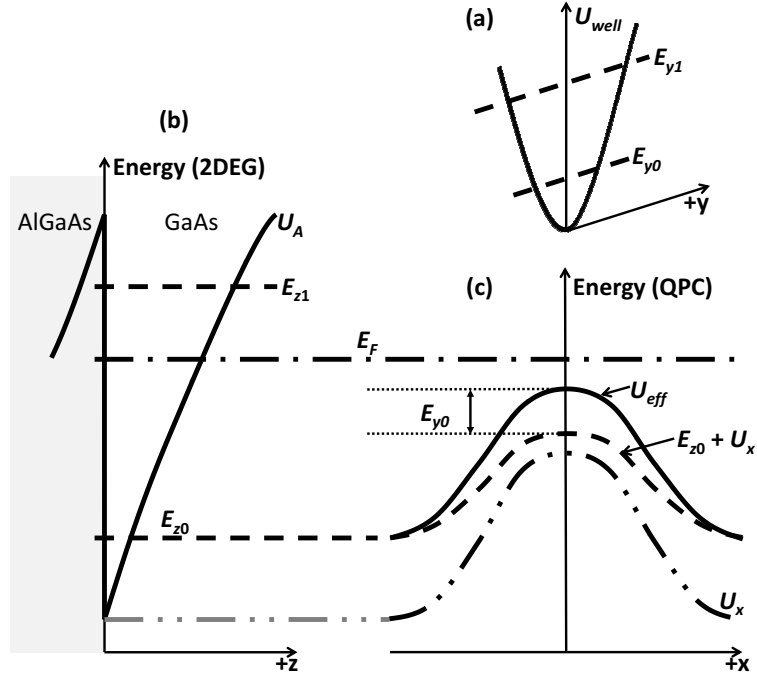}

\caption{\label{fig:2-potential}(a) Confinement potential well $U_{\mathrm{well}}$
induced by the gates, showing the energy sub-levels $E_{y0}$, $E_{y1}$,
etc. (b) Potential in the 2DEG, showing the Fermi energy and the 2DEG
subbands. (c) Contributions to the effective potential: energy of
the lowest subband ($E_{z0}$), potential barrier ($U_{x}$) and energy
due to the parabolic confinement ($E_{y0}$).}
\end{figure*}

A top view of the transport configuration is shown schematically in
Figure \ref{fig:1-scheme}. The $x$-direction runs between the source
and drain electrodes, while the $y$-axis extends from one gate to
the other. The center of the QPC is at $\left(x,y\right)=\left(0,0\right)$.
Electrons are injected into the QPC from the 2DEG on the source side
of the structure and collected at the drain. As a consequence of the
finite width of the negatively-biased gate electrodes, there is a
saddle potential $U_{\mathrm{QPC}}\left(x,y\right)$ in the QPC that
results from two contributions: first, the potential energy well $U_{\mathrm{well}}$,
which we model as parabolic (see Fig. \ref{fig:2-potential}(a)),
confines electrons to a narrow channel between the gates along the
$y$-direction, and splits each of the 2DEG subbands (with energies
$E_{z0}$, $E_{z1}$,... away from the QPC for $x\rightarrow-\infty$)
into a new set of discrete 1D energy subbands with energies $E_{y0}$,
$E_{y1}$,... above those of the 2DEG subbands. Second, there is also
a smooth potential barrier $U_{x}$ running along the $x$-direction.
Due to the saddle potential, the charge density at the QPC decreases,
which in turn reduces the effect of electron-electron interactions
and shifts the 2DEG subband energies downwards. The addition of the
energy $E_{z0}$ of the lowest 2DEG subband (Fig. \ref{fig:2-potential}(b)),
the (smallest) energy increase $E_{y0}$ from the parabolic confinement,
and the QPC potential barrier $U_{x}$, results in an effective 1D
potential $U_{\mathrm{eff}}$, as shown in Fig. \ref{fig:2-potential}(c).

In our study we limit ourselves to the lowest-energy subband and include
the Coulomb and exchange interactions ($U_{\mathrm{el}}$ and $U_{\mathrm{exch}}$,
respectively) within the Hartree-Fock approximation. The Fermi energy
$E_{F}$ is pinned at the source electrode, whereas the Fermi level
at the drain is shifted from this value by a small applied source-drain
bias. We neglect the spin-orbit (Rashba) interaction $\hat{H}_{so}=\alpha\left(\hat{\boldsymbol{\sigma}}\times\hat{\mathbf{p}}\right)_{z}/\hbar$
, as its contribution to the effective potential in GaAs will be much
smaller than in other materials. The Rashba coupling constant in GaAs
is $\alpha=\left(\unit[0.052]{e\cdot nm^{2}}\right)F_{z}$ (where
$F_{z}$ is the electric field in the $z$-direction), which is significantly
less than in InAs ($\left(\unit[1.17]{e\cdot nm^{2}}\right)F_{z}$)
and in InSb ($\left(\unit[5.23]{e\cdot nm^{2}}\right)F_{z}$)\citealp{Winkler2003}.
We also neglect the image potential, considering the relatively small
difference between the dielectric constants of GaAs ($\kappa=12.9$)
and $\mathrm{Al_{x}Ga_{1-x}As}$ ($\kappa=12.0$ for $x=0.3$).

\section{3D Hartree-Fock Model of Modulation-Doped 1D GaAs Channels}

Before solving for the electronic properties in the QPC we consider
the case of a very long constriction, effectively a 1D channel or
quantum wire. In this case, the system is translationally invariant
along the $x$-direction and the QPC potential barrier $U_{x}$ in
the wire is reduced to a constant value $U_{0}$.

In the absence of an applied split-gate voltage, the (2DEG) confinement
potential in GaAs due to the depletion from ionized acceptors is well
approximated by\citealp{Ando1982} $U_{\mathrm{A}}\left(z\right)=\frac{q^{2}N_{A}}{\epsilon}z\left(z_{d}-\frac{z}{2}\right)$;
here, $q$ is the electron charge, $\epsilon$ is the GaAs dielectric
constant, $N_{A}$ is the acceptor density and $z_{d}=\sqrt{2\epsilon E_{C}/q^{2}N_{A}}$
is the width of the depletion region ($E_{C}$ being the bottom of
the conduction band in the bulk of GaAs). This potential satisfies
the boundary conditions $\left[\partial U_{A}/\partial z\right]\left(z=z_{d}\right)=0$
(i.e . the electric field is zero at $z=z_{d}$) and $U_{A}\left(z=0\right)=0$.
The parabolic confining potential induced by the applied gate voltage
is given by $U_{\mathrm{well}}\left(y\right)=m^{*}\omega_{y}^{2}y^{2}/2$,
where $\omega_{y}$ represents the strength of the confinement.

With these confining potentials, the Schrödinger equation within the
Hartree-Fock approximation reads
\begin{multline}
-\frac{\hbar^{2}}{2m^{*}}\nabla^{2}\psi_{\left\{ i\right\} }\left(\vec{r}\right)+\left[U_{A}\left(z\right)+U_{\mathrm{well}}\left(y\right)+U_{x}+\right.\\
\left.U_{\mathrm{el}}\left(\vec{r}\right)\right]\psi_{\left\{ i\right\} }\left(\vec{r}\right)+\hat{U}_{\mathrm{exch}}\left[\psi_{\left\{ j\right\} }\left(\vec{r}\right)\right]=E_{\left\{ i\right\} }\psi_{\left\{ i\right\} }\left(\vec{r}\right)\label{eq:Schrodinger1}
\end{multline}
where the quantum numbers $\left\{ i\right\} =\left\{ i_{x},i_{y},i_{z},\sigma_{i}\right\} $
($\sigma$ being the electron spin) are associated with the eigenenergies
$E_{\left\{ i\right\} }$.
\begin{widetext}
The Hartree (direct) term $U_{\mathrm{el}}$ and the exchange term
$\hat{U}_{\mathrm{exch}}$ read, respectively:
\begin{eqnarray}
U_{\mathrm{el}}\left(\vec{r}\right)\psi_{\left\{ i\right\} }\left(\vec{r}\right) & = & \hphantom{\textnormal{-}}\sum_{\left\{ j\right\} }\int\mathrm{d}^{3}\vec{r}^{\,\prime}\left|\psi_{\left\{ j\right\} }\left(\vec{r}^{\,\prime}\right)\right|^{2}U_{\mathrm{Coul}}\left(\vec{r},\vec{r}^{\,\prime}\right)\psi_{\left\{ i\right\} }\left(\vec{r}\right)\\
\hat{U}_{\mathrm{exch}}\left[\psi_{\left\{ j\right\} }\left(\vec{r}\right)\right] & = & \textnormal{-}\sum_{\left\{ j\right\} }\int\mathrm{d}^{3}\vec{r}^{\,\prime}\,\psi_{\left\{ j\right\} }^{*}\left(\vec{r}^{\,\prime}\right)U_{\mathrm{Coul}}\left(\vec{r},\vec{r}^{\,\prime}\right)\psi_{\left\{ i\right\} }\left(\vec{r}^{\,\prime}\right)\psi_{\left\{ j\right\} }\left(\vec{r}\right)\delta_{\sigma_{i}\sigma_{j}}
\end{eqnarray}

$U_{\mathrm{Coul}}$, the Coulomb interaction, is given by
\begin{equation}
U_{\mathrm{Coul}}\left(\vec{r},\vec{r}^{\,\prime}\right)=\frac{q^{2}}{4\pi\epsilon}\left(\frac{1}{\left|\vec{r}^{\,\prime}-\vec{r}\right|}-\frac{1}{\sqrt{\left(x^{\prime}-x\right)^{2}+\left(y^{\prime}-y\right)^{2}+\left(z^{\prime}+z\right)^{2}}}\right)\label{eq:Ucoul}
\end{equation}
The second term on the right-hand side corresponds to mirror charges
placed on the AlGaAs side of the interface. With this expression for
$U_{\mathrm{Coul}}$, $U_{\mathrm{el}}\left(\vec{r}\right)$ satisfies
boundary conditions similar to those of $U_{A}$, i.e. $U_{\mathrm{el}}\left(z=0\right)=0$
and $\left[\partial U_{\mathrm{el}}/\partial z\right]_{z=z_{d}}\approx\left[\partial U_{\mathrm{el}}/\partial z\right]_{z\rightarrow\infty}=0$.
(This will also ensure that the expectation value $\left\langle U_{\mathrm{el}}\right\rangle $
remains finite.) The occupation number of a state with energy $E$
is given by the Fermi-Dirac distribution, $f_{T}\left(E\right)=\left\{ 1+\exp\left[\left(E-\mu\right)/k_{B}T\right]\right\} ^{-1}$,
where $\mu$ is the chemical potential (which equals $E_{F}$ at $T=0$).

Since $U_{x}\left(=U_{0}\right)$ is constant, the solution to Eq.
\eqref{eq:Schrodinger1} can be written as a product of a plane wave
traveling along $x$ and a function depending on $y$ and $z$, i.e.
$\psi_{\left\{ k_{x},i_{y},i_{z},\sigma_{i}\right\} }=\left(\mathrm{e}^{\mathrm{i}k_{x}x}/\sqrt{L_{x}}\right)\Lambda_{i_{y},i_{z}}\left(y,z\right)$,
where $L_{x}$ is the length of the wire. With this assumption, the
$y$- and $z$-dependence in Eq. \eqref{eq:Schrodinger1} can be separated,
yielding
\begin{multline}
\left\{ -\frac{\hbar^{2}}{2m^{*}}\left(\partial_{y}^{2}+\partial_{z}^{2}\right)U_{A}\left(z\right)+U_{\mathrm{well}}\left(y\right)+U_{x}+U_{\mathrm{el}}\left(y,z\right)\right\} \Lambda_{i_{y},i_{z}}\left(y,z\right)\\
+\hat{U}_{\mathrm{exch}}\left[\Lambda_{j_{y},j_{z}}\left(y,z;k_{x},\sigma_{j}\right)\right]=\left\{ E\left(k_{x},i_{y},i_{z},\sigma_{i}\right)-\frac{\hbar^{2}k_{x}^{2}}{2m^{*}}\right\} \Lambda_{i_{y},i_{z}}\left(y,z\right)\label{eq:Sch-separated}
\end{multline}
where, after expanding $U_{\mathrm{Coul}}$ (Eq. \eqref{eq:Ucoul})
in a Fourier series, the Hartree and exchange terms are respectively
given by
\begin{equation}
U_{\mathrm{el}}\left(y,z\right)=\sum_{p_{x},j_{y},j_{z},\sigma_{j}}\int\mathrm{d}y^{\prime}\int\mathrm{d}z^{\prime}\left|\Lambda_{j_{y},j_{z}}\left(y^{\prime},z^{\prime}\right)\right|^{2}\frac{1}{L_{x}}\int\frac{\mathrm{d}p_{y}}{p_{y}}\mathrm{e}^{\mathrm{i}p_{y}\left(y^{\prime}-y\right)}\left[\mathrm{e}^{-p_{y}\left|z^{\prime}-z\right|}-\mathrm{e}^{-p_{y}\left(z^{\prime}+z\right)}\right]
\end{equation}
\begin{eqnarray}
\hat{U}_{\mathrm{exch}}\left[\Lambda_{j_{y},j_{z}}\left(y,z;k_{x},\sigma_{j}\right)\right] & = & \textnormal{-}\hspace{-1em}\sum_{p_{x},j_{y},j_{z},\sigma_{j}}\delta_{\sigma_{i}\sigma_{j}}\int\mathrm{d}y^{\prime}\int\mathrm{d}z^{\prime}\,\Lambda_{j_{y},j_{z}}^{*}\left(y^{\prime},z^{\prime}\right)\Lambda_{i_{y},i_{z}}\left(y^{\prime},z^{\prime}\right)\Lambda_{j_{y},j_{z}}\left(y,z\right)\nonumber \\
 &  & \cdot\frac{1}{L_{x}}\int\frac{\mathrm{d}p_{y}}{\bar{p}_{y}}\mathrm{e}^{\mathrm{i}p_{y}\left(y^{\prime}-y\right)}\left[\mathrm{e}^{-\bar{p}_{y}\left|z^{\prime}-z\right|}-\mathrm{e}^{-\bar{p}_{y}\left(z^{\prime}+z\right)}\right]
\end{eqnarray}
with $\bar{p}_{y}\equiv\sqrt{\left(p_{x}-k_{x}\right)^{2}+p_{y}^{2}}$.
\end{widetext}
We take the expectation value of the left-hand side of Eq. \eqref{eq:Sch-separated}
and we define two new potential energies: $E_{yz}$, which corresponds
to the carrier kinetic energy and the confinement along the y- and
z-directions, and $U_{\mathrm{ee}}$, which corresponds to electron-electron
interactions:
\begin{equation}
E_{yz}\equiv\left\langle -\frac{\hbar^{2}}{2m^{*}}\left(\partial_{y}^{2}+\partial_{z}^{2}\right)\right\rangle +\left\langle U_{\mathrm{well}}\right\rangle +\left\langle U_{A}\right\rangle 
\end{equation}
\begin{equation}
U_{\mathrm{ee}}\equiv\left\langle U_{\mathrm{el}}\right\rangle +\left\langle \hat{U}_{\mathrm{exch}}\right\rangle 
\end{equation}

Then, we define the effective 1D potential as
\begin{multline}
U_{\mathrm{eff}}\left(k_{x},i_{y},i_{z},\sigma_{i}\right)=\\
U_{x}+E_{yz}\left(i_{y},i_{z}\right)+U_{\mathrm{ee}}\left(k_{x},i_{y},i_{z},\sigma_{i}\right)\label{eq:Ueff-x}
\end{multline}
so that the 1D electron energy reads
\begin{equation}
E\left(k_{x},i_{y},i_{z},\sigma_{i}\right)=\frac{\hbar^{2}k_{x}^{2}}{2m^{*}}+U_{\mathrm{eff}}\left(k_{x},i_{y},i_{z},\sigma_{i}\right)\label{eq:E-kx}
\end{equation}

To study the extreme quantum limit, when only the lowest-energy subband
is occupied, we use the trial wave function defined as
\begin{equation}
\Lambda_{00}\left(y,z\right)=\left(\frac{a^{\nicefrac{1}{2}}}{\pi^{\nicefrac{1}{4}}}\mathrm{e}^{-a^{2}y^{2}/2}\right)\left(\frac{b^{\nicefrac{3}{2}}}{2^{\nicefrac{1}{2}}}z\mathrm{e}^{-bz/2}\right)
\end{equation}
where the parameters $a$ and $b$ are determined by the variational
method. The first term of the right-hand side is the ground state
of the parabolic potential $U_{\mathrm{well}}$, while the second
term approximates the ground state of the depletion potential $U_{A}$.
\begin{widetext}
Then, we obtain the following expectation values for the different
terms in the effective potential $U_{\mathrm{eff}}$:\begin{subequations}
\begin{equation}
\left\langle -\frac{\hbar^{2}}{2m^{*}}\left(\partial_{y}^{2}+\partial_{z}^{2}\right)\right\rangle \equiv\left\langle \hat{T}_{yz}\right\rangle =\frac{\hbar^{2}}{8m^{*}}\left(2a^{2}+b^{2}\right)\label{eq:Tyz}
\end{equation}
\begin{equation}
\left\langle U_{\mathrm{well}}\right\rangle +\left\langle U_{A}\right\rangle =\frac{m^{*}\omega_{y}^{2}}{4a^{2}}+\frac{3q^{2}N_{A}z_{d}}{\epsilon b}\left(1-\frac{2}{z_{d}b}\right)\label{eq:Uw-p-UA}
\end{equation}
\begin{equation}
\left\langle U_{\mathrm{el}}\right\rangle =\frac{q^{2}}{16\pi^{2}\epsilon}\zeta_{ab}\left(0\right)\int_{-\infty}^{+\infty}\mathrm{d}p_{x}\, f_{T}\left[E\left(p_{x}\right)\right]=\frac{q^{2}\zeta_{ab}\left(0\right)}{16\pi\epsilon}n_{0}\label{eq:Uel}
\end{equation}
\begin{equation}
\left\langle \hat{U}_{\mathrm{exch}}\left(k_{x}\right)\right\rangle =-\frac{q^{2}}{32\pi^{2}\epsilon}\int_{-\infty}^{+\infty}\mathrm{d}p_{x}\,\zeta_{ab}\left(p_{x}-k_{x}\right)f_{T}\left[E\left(p_{x}\right)\right]\label{eq:Uexch}
\end{equation}

\end{subequations}

Here, $n_{0}=\frac{1}{\pi}\int_{-\infty}^{+\infty}\mathrm{d}k\, f_{T}\left(E\left(k\right)\right)$
is the 1D electron density in the wire and $\zeta_{ab}$ is a dimensionless
function given by
\begin{equation}
\zeta_{ab}\left(p_{x}-k_{x}\right)=b\int_{0}^{\infty}\mathrm{d}p_{y}\, e^{-p_{y}^{2}/2a^{2}}\frac{3\bar{p}_{y}^{4}+18b\bar{p}_{y}^{3}+44b^{2}\bar{p}_{y}^{2}+54b^{3}\bar{p}_{y}+33b^{4}}{\left(\bar{p}_{y}+b\right)^{6}}
\end{equation}

Parameters $a$ and $b$ are obtained by minimizing the average effective
energy per electron, $U_{\mathrm{avg}}=E_{yz}+\frac{1}{2}\left\langle U_{\mathrm{el}}\right\rangle +J_{\mathrm{avg}}$,
where $J_{\mathrm{avg}}$ is the average exchange energy per electron:
\begin{equation}
J_{\mathrm{avg}}=-\frac{q^{2}}{64\pi^{3}\epsilon n_{0}}\int_{-\infty}^{+\infty}\mathrm{d}k_{x}\, f_{T}\left[E\left(k_{x}\right)\right]\int_{-\infty}^{+\infty}\mathrm{d}p_{x}\, f_{T}\left[E\left(p_{x}\right)\right]\zeta_{ab}\left(p_{x}-k_{x}\right)
\end{equation}

\end{widetext}
As a way to check the validity of this model, we notice that the expressions
for $\left\langle -\frac{\hbar^{2}}{2m^{*}}\partial_{z}^{2}\right\rangle \equiv\left\langle \hat{T}_{z}\right\rangle $
and$\left\langle U_{A}\right\rangle $ (Eq. \eqref{eq:Tyz}-\eqref{eq:Uw-p-UA}),
in the zero-confinement case ($\omega_{y}=0$) reduce to the corresponding
expressions for the 2DEG, which are given by $\left\langle \hat{T}_{z\mathrm{,2D}}\right\rangle =\frac{\hbar^{2}b^{2}}{8m^{*}}$
and $\left\langle U_{A\mathrm{,2D}}\right\rangle =\frac{3q^{2}N_{A}z_{d}}{\epsilon b}\left(1-\frac{2}{bz_{d}}\right)$.
We also compare $\left\langle U_{\mathrm{el}}\right\rangle $ (Eq.
\eqref{eq:Uel}) to its 2DEG counterpart\citealp{Ando1982}, $\left\langle U_{\mathrm{el,2D}}\right\rangle =\frac{33q^{2}n_{2D}}{16\epsilon b}$,
by first expanding $\zeta_{ab}\left(0\right)$ in a series:
\begin{equation}
\zeta_{ab}\left(0\right)=\frac{33\sqrt{\pi}}{\sqrt{2}}\frac{a}{b}-144\left(\frac{a}{b}\right)^{2}+\frac{413\sqrt{\pi}}{\sqrt{2}}\left(\frac{a}{b}\right)^{3}+...
\end{equation}

Then, to first order in $a/b$, $\left\langle U_{\mathrm{el}}\right\rangle =\frac{33q^{2}n_{0}a}{16\sqrt{2\pi}\epsilon b}$,
which matches $\left\langle U_{\mathrm{el,2D}}\right\rangle $ if
we identify $n_{2D}$ with $\frac{n_{0}a}{\sqrt{2\pi}}$.

In general, the effective potential $U_{\mathrm{eff}}\left(k_{x}\right)$
and the electron density $n_{0}$ are obtained self-consistently by
solving the integral equation for $E\left(k_{x}\right)$ (Eq. \eqref{eq:E-kx})
numerically. However, we note that if the exchange term is negligible
compared to the Hartree term (which is the case when $n_{0}$ is sufficiently
large), an explicit solution for $n_{0}$ can be obtained at $T=0$:
\begin{widetext}
\begin{eqnarray}
n_{0}\left[T=0\right] & = & 2\sqrt{\frac{2m^{*}}{\pi^{2}\hbar^{2}}}\,\theta\left(E_{F}-U_{x}-E_{yz}\right)\nonumber \\
 &  & \times\left\{ -\sqrt{\frac{2m^{*}}{\pi^{2}\hbar^{2}}}\left(\frac{q^{2}\zeta_{ab}\left(0\right)}{16\pi\epsilon}\right)+\sqrt{\frac{2m^{*}}{\pi^{2}\hbar^{2}}\left(\frac{q^{2}\zeta_{ab}\left(0\right)}{16\pi\epsilon}\right)^{2}+\left(E_{F}-U_{x}-E_{yz}\right)}\right\} \label{eq:n0-0K}
\end{eqnarray}
For $T>0$, the solution is obtained by solving the following expression
numerically:
\begin{equation}
n_{0}\left[T\right]=\sqrt{\frac{2m^{*}k_{B}T}{\pi^{2}\hbar^{2}}}F_{-1/2}\left[\frac{1}{k_{B}T}\left(\mu-U_{x}-E_{yz}-\frac{q^{2}\zeta_{ab}\left(0\right)}{16\pi\epsilon}n_{0}\right)\right]
\end{equation}

\end{widetext}
Here, $F_{-1/2}$ is the Fermi-Dirac integral of order $-1/2$.

\begin{figure}
\noindent \includegraphics[width=2.8in]{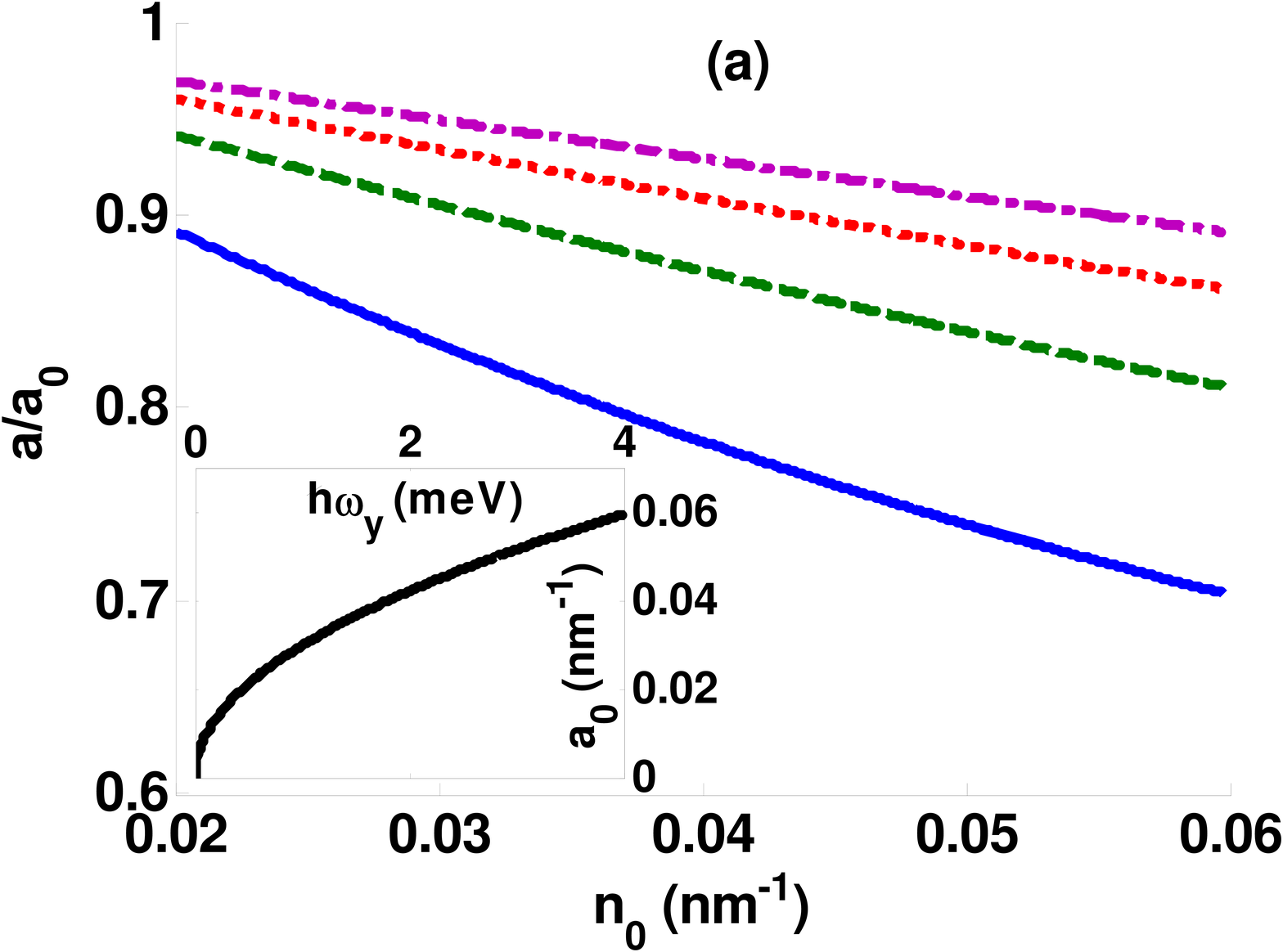}

\noindent \includegraphics[width=2.8in]{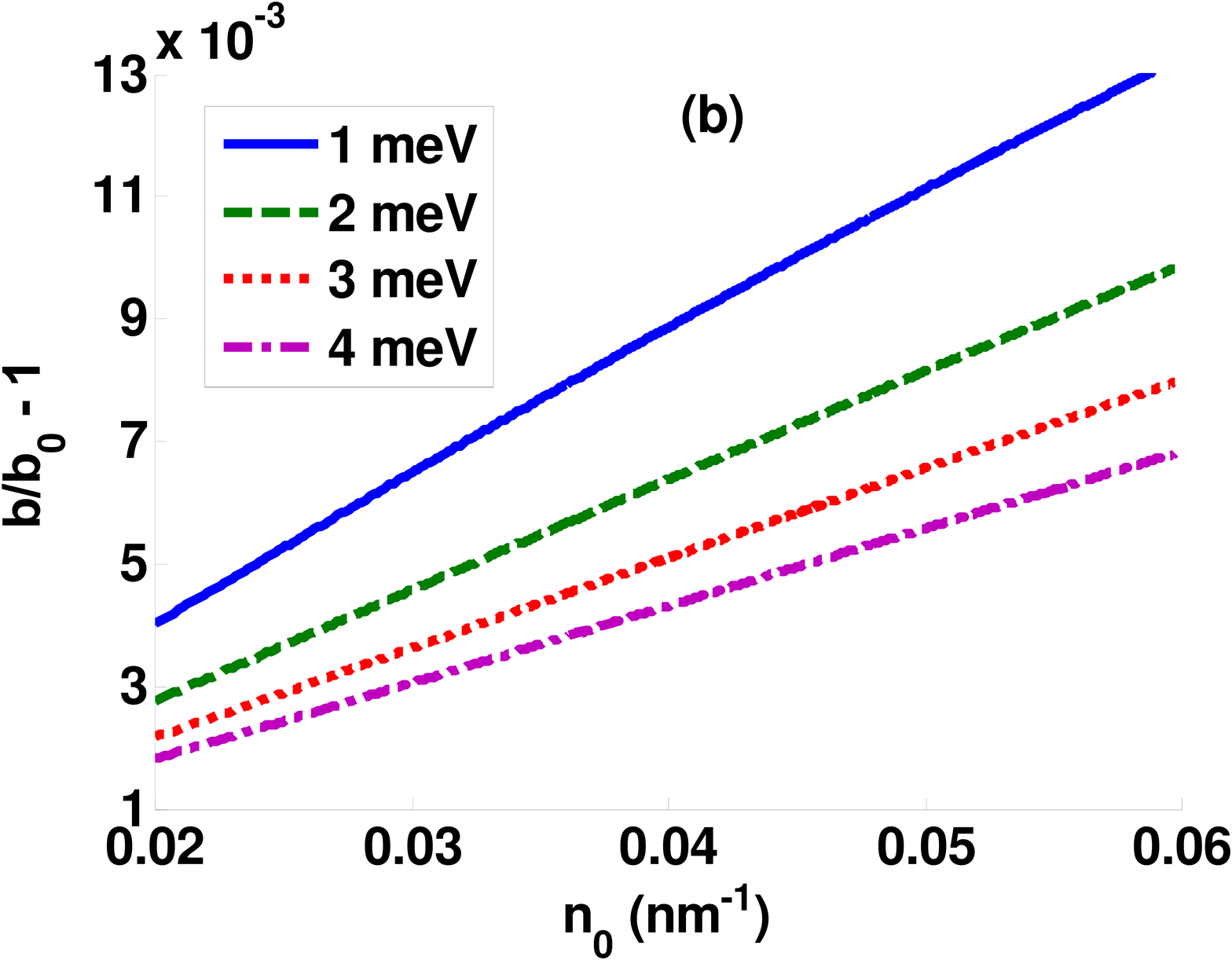}

\caption{\label{fig:3-ab-vs-n0-omegay}(Color online) (a) $a/a_{0}$ and (b)
$b/b_{0}-1$ vs. electron density $n_{0}$ for different confinement
strengths $\hbar\omega_{y}$ in the wire, from $\unit[1]{meV}$ (solid
line) to $\unit[4]{meV}$ (dot-dashed line), for $N_{A}=\unit[10^{14}]{cm^{-3}}$,
at $T=0$. Inset in (a): $a_{0}$ vs. $\hbar\omega_{y}$. For (b),
$b_{0}=\unit[0.190]{nm^{-1}}$.}
\end{figure}

Meanwhile, if exchange effects are included, the zero-temperature
solution for $n_{0}$ (written in terms of $k_{F}=\pi n_{0}/2$) is
given by the solution of:
\begin{multline}
E_{F}=\frac{\hbar^{2}k_{F}^{2}}{2m^{*}}+U_{\mathrm{eff}}=\frac{\hbar^{2}k_{F}^{2}}{2m^{*}}+U_{x}+E_{yz}\\
+\frac{q^{2}}{32\pi^{2}\epsilon}\left[4k_{F}\zeta_{ab}\left(0\right)-\int_{0}^{2k_{F}}\mathrm{d}k_{x}\,\zeta_{ab}\left(k_{x}\right)\right]\label{eq:kF-eqn-exch-T0}
\end{multline}

For very small $k_{F}$, the integral in Eq. \eqref{eq:kF-eqn-exch-T0}
above is approximately equal to $2k_{F}\zeta_{ab}\left(0\right)$,
since $\zeta_{ab}\left(k\right)=\zeta_{ab}\left(0\right)+\mathrm{O}\left(k^{2}\right)$.
Therefore, in this situation, $\left\langle \hat{U}_{\mathrm{exch}}\right\rangle \approx-\frac{1}{2}\left\langle U_{\mathrm{el}}\right\rangle $
and Eq. \eqref{eq:n0-0K} can be used to obtain an approximate solution
for $n_{0}$ in the presence of the exchange interaction, provided
that $\zeta_{ab}\left(0\right)$ is replaced with $\zeta_{ab}\left(0\right)/2$.

Figures \ref{fig:3-ab-vs-n0-omegay} (a) and (b) display the variational
parameters $a$ and $b$ (normalized to their values when $n_{0}=0$,
$a_{0}=\sqrt{m^{*}\omega_{y}/\hbar}$ and $b_{0}\approx\left[12m^{*}q^{2}N_{A}z_{d}/\hbar^{2}\epsilon\right]^{1/3}$)
versus the electron density $n_{0}$ in the wire for different confinement
strengths $\hbar\omega_{y}$. Calculations are carried out using the
parameters for GaAs\citealp{Hess2000}: $\epsilon=12.9\epsilon_{0}$,
$m^{*}=0.067m_{0}$, $E_{C}=\unit[1.52]{eV}$. For a wire length of
a few hundred nanometers, the calculated electron densities correspond
to a population that varies from a few electrons to tens of electrons
in the wire.

Figure \ref{fig:3-ab-vs-n0-omegay} shows that, as $n_{0}$ increases,
$a$ decreases relative to $a_{0}$ (and the characteristic length
of the wavefunction along $y$, which is proportional to $1/a$, increases).
This indicates that electron-electron interactions counteract the
effects of the gate-induced lateral confinement and spread out the
electrons over a wider region along the $y$-axis. $a_{0}$ itself
increases with increasing confinement strength, as seen in the inset
of Fig. \ref{fig:3-ab-vs-n0-omegay}(a). Meanwhile, $b$ increases
relative to $b_{0}$ when $n_{0}$ increases, as seen in Figure \ref{fig:3-ab-vs-n0-omegay}(b),
which means that electrons are confined to a narrower region near
the heterojunction interface along the $z$-axis. The variation in
both parameters relative to their zero-population values is more pronounced
for smaller $\hbar\omega_{y}$, but $a$ is more sensitive to changes
in the electron density than $b$, e.g. for $n_{0}=\unit[5\times10^{5}]{cm^{-1}}$,
Figures \ref{fig:3-ab-vs-n0-omegay}(a) and (b) show that $a$ decreases
by up to 25\% (for $\hbar\omega_{y}=\unit[1]{meV}$), while $b$ goes
up by 1\% at most for the same confinement strength. The small variation
of $b$, coupled with the fact that $1/b$ (the characteristic length
of the wavefunction along $z$) is of the order of a few nanometers
and an order of magnitude smaller than $1/a$, is consistent with
the quasi-2DEG nature of the electron layer in GaAs.

\begin{figure}
\noindent \includegraphics[width=2.8in]{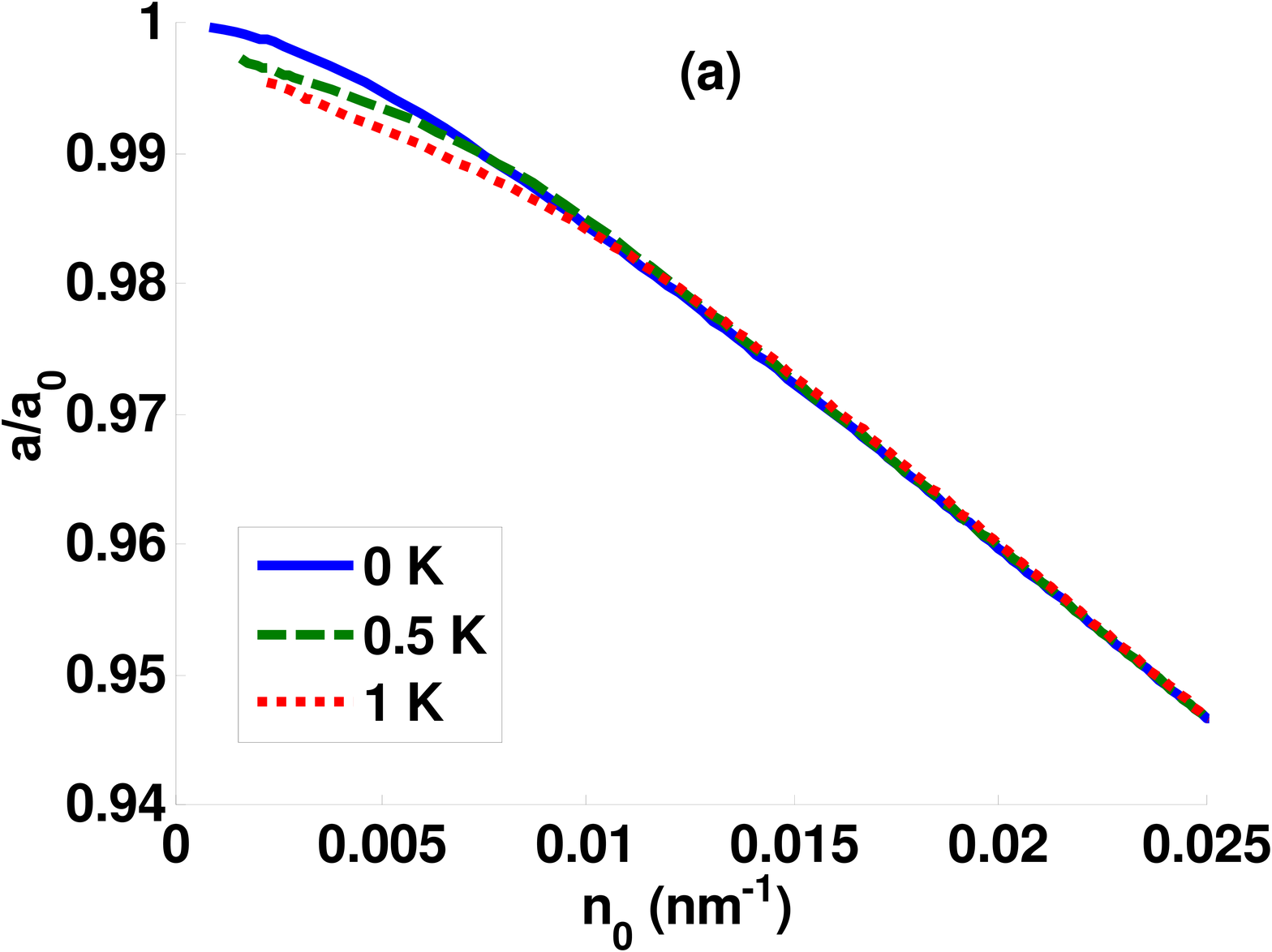}

\noindent \includegraphics[width=2.8in]{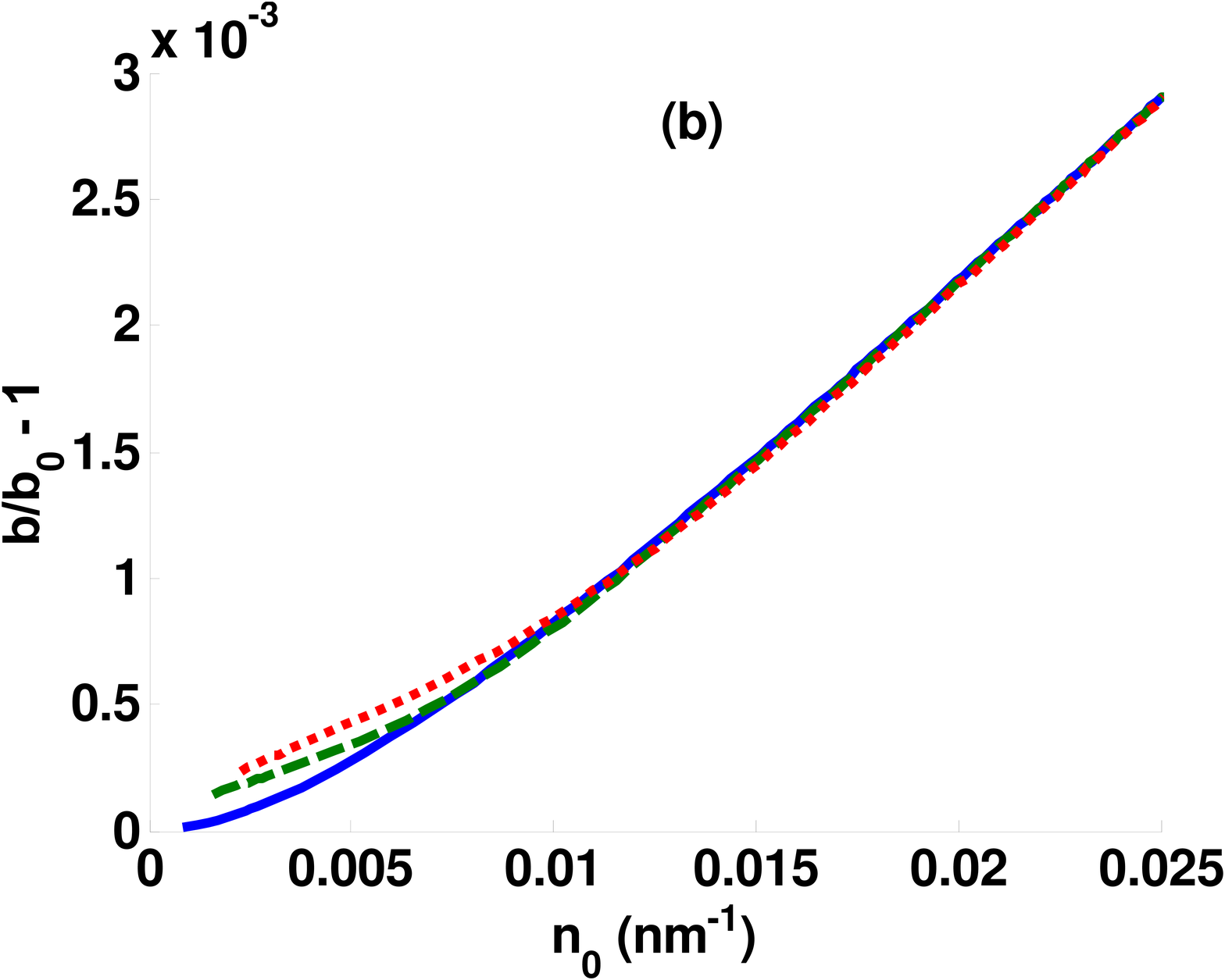}

\caption{\label{fig:4-ab-vs-n0-Temp}(Color online) (a) $a/a_{0}$ and (b)
$b/b_{0}$ vs. $n_{0}$ for different temperatures, with $\hbar\omega_{y}=\unit[2]{meV}$
and $N_{A}=\unit[10^{14}]{cm^{-3}}$. $a_{0}=\unit[0.0419]{nm^{-1}}$
and $b_{0}=\unit[0.190]{nm^{-1}}$.}
\end{figure}

\begin{figure}
\noindent \includegraphics[width=2.8in]{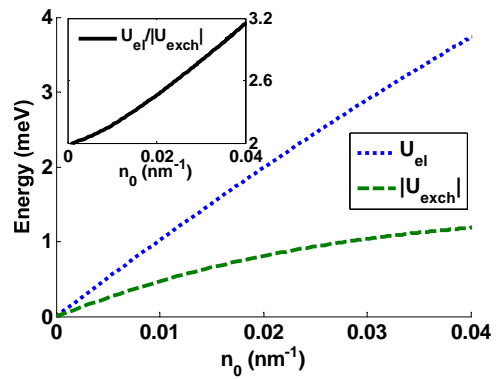}

\caption{\label{fig:5-dir-exch-vs-n0}(Color online) Hartree and exchange terms,
$U_{\mathrm{el}}$ and $U_{\mathrm{exch}}$, versus $n_{0}$ in a
wire at $T=0$. $\hbar\omega_{y}=\unit[2]{meV}$ and $N_{A}=\unit[10^{14}]{cm^{-3}}$.
Inset: $U_{\mathrm{el}}/\left|U_{\mathrm{exch}}\right|$ vs. $n_{0}$.}
\end{figure}

Figures \ref{fig:4-ab-vs-n0-Temp} (a) and (b) show the temperature
dependence of $a$ and $b$. Even at $\unit[1]{K}$, $a$ and $b$
do not vary significantly when compared to their zero-temperature
values. The largest variations occur for very low electron densities,
with $a$ varying by less than 0.5\% (respect to its zero-temperature
value) and $b$ by less than 0.05\%. For high electron densities,
the effect of temperature on the variational parameters is negligible.

Figure \ref{fig:5-dir-exch-vs-n0} displays the variation of the Hartree
and exchange terms versus the electron density $n_{0}$ in the wire.
As suggested by Eq. \eqref{eq:Uel}, the Hartree term $U_{\mathrm{el}}$
grows approximately linearly with respect to $n_{0}$. Deviations
from linearity are due to the fact that both $a$ and $b$ depend
on $n_{0}$. The exchange term $\left|U_{\mathrm{exch}}\right|$ grows
at a slower rate than $U_{\mathrm{el}}$ with increasing $n_{0}$.
For low concentrations, $\left|U_{\mathrm{exch}}\right|$ is equal
to one-half of $U_{\mathrm{el}}$, as predicted by Eq. \eqref{eq:kF-eqn-exch-T0}.

\section{Transport model for the QPC}

When these results are extended to a QPC of finite length, $U_{x}$
does not take a constant value anymore, but rather depends on $x$.
Additionally, the lateral confinement $U_{\mathrm{well}}$ will now
drop to zero as $x$ increases. Thus, the Schrödinger equation (Eq.
\eqref{eq:Schrodinger1}) is no longer separable. However, if $U_{x}$
varies smoothly and slowly relative to $U_{\mathrm{well}}$ and $U_{A}$,
an adiabatic approximation can be carried out in Eq. \eqref{eq:Schrodinger1}
to find local energy levels $E\left(k_{x};x\right)$, a local electron
density $n_{0}\left(x\right)$ and $x$-dependent parameters $a$
and $b$.

\begin{figure*}
\noindent \includegraphics[width=3.1in]{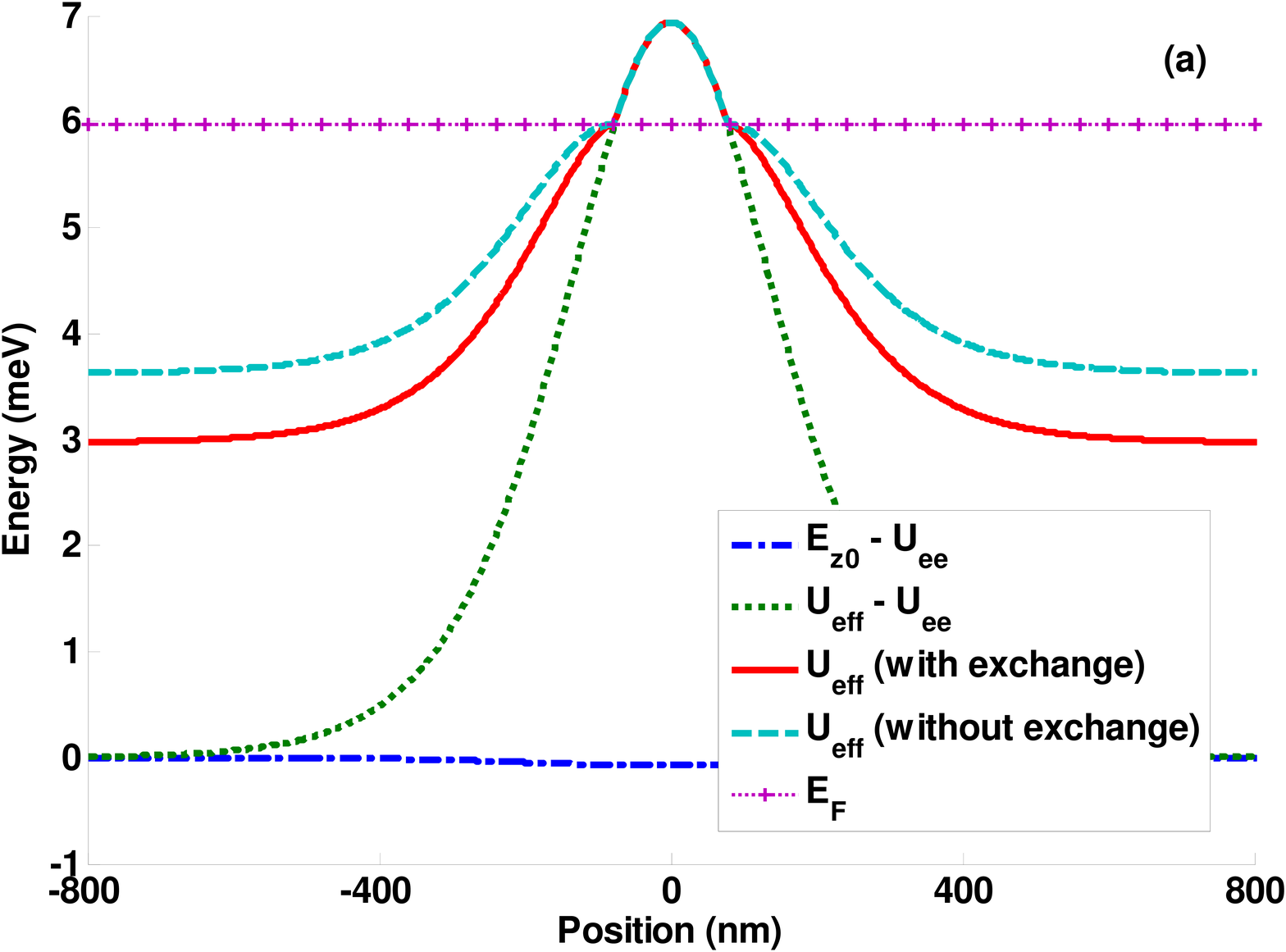}\includegraphics[width=3.1in]{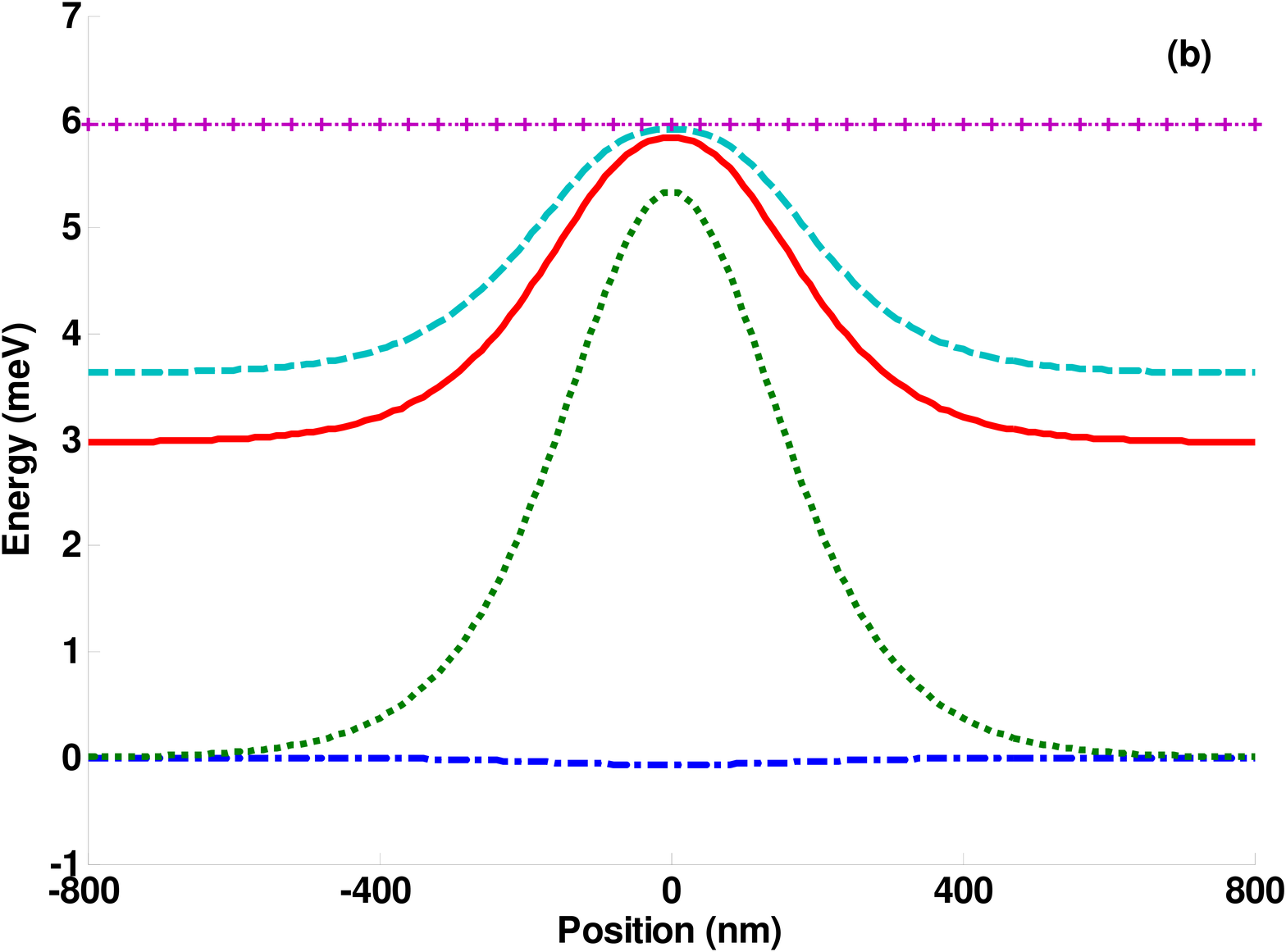}

\noindent \includegraphics[width=3.1in]{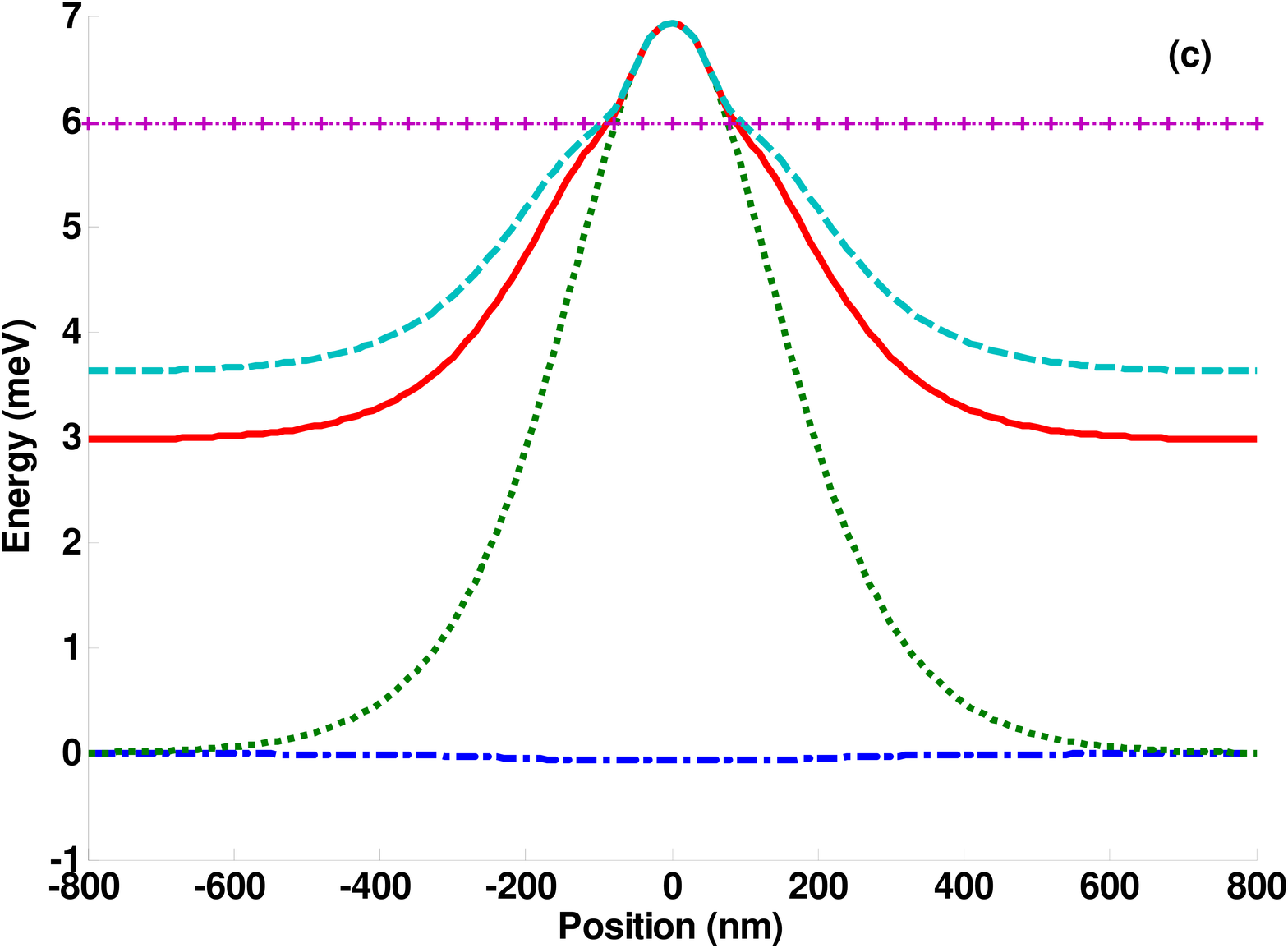}\includegraphics[width=3.1in]{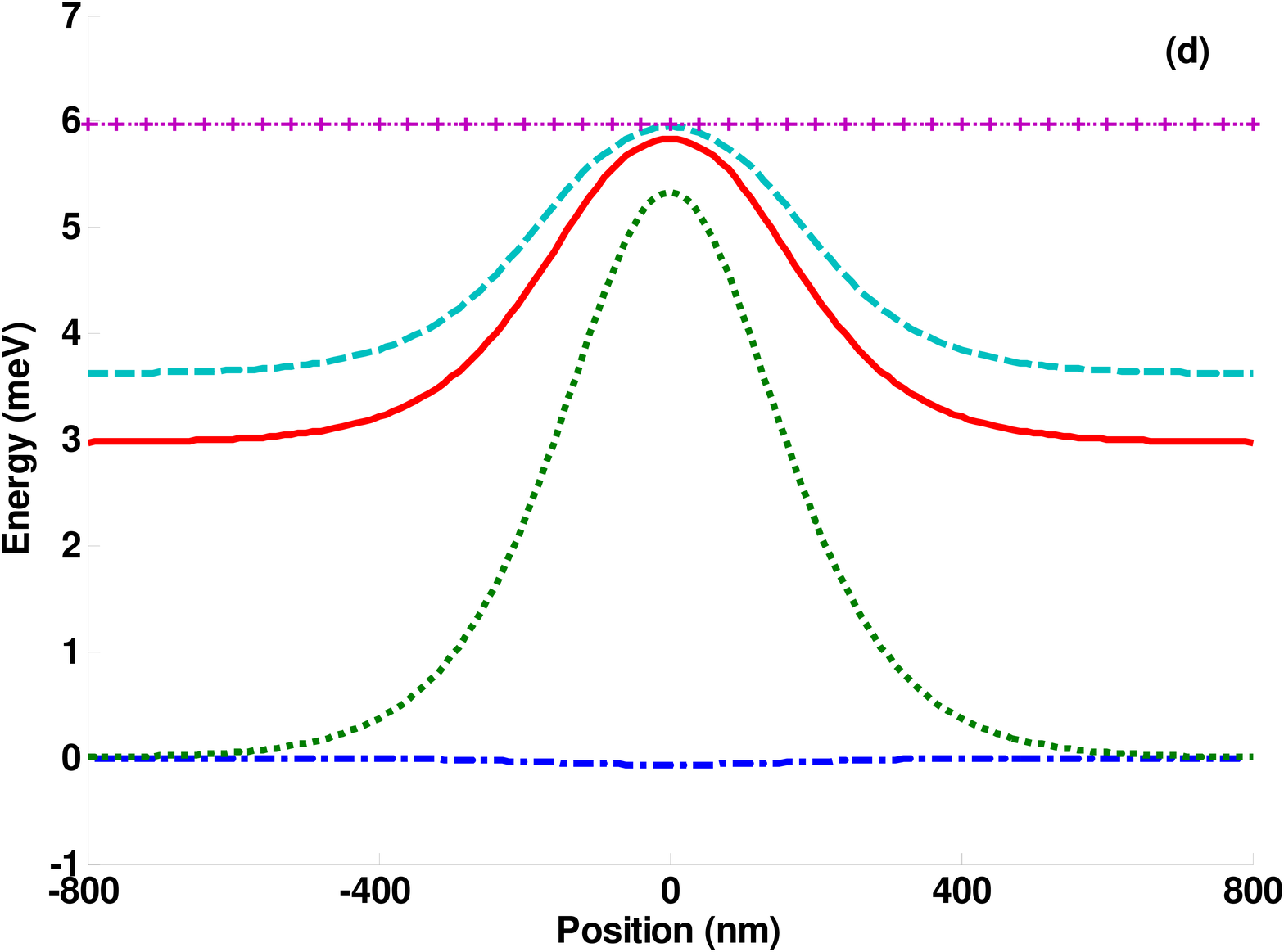}

\caption{\label{fig:6-Ueff-profile-0K}(Color online) Effective potential vs.
position when: (a), (c) $U_{\mathrm{eff,max}}>E_{F}$ ($U_{0}=\unit[7]{meV}$);
(b), (d) $U_{\mathrm{eff,max}}<E_{F}$ ($U_{0}=\unit[5.4]{meV}$).
For (a) and (b), $T=\unit[0]{K}$; for (c) and (d), $T=\unit[1]{K}$.
$\hbar\omega_{y,\mathrm{max}}=\unit[2]{meV}$, $N_{A}=\unit[10^{14}]{cm^{-3}}$,
$n_{2D}=\unit[8.4\times10^{10}]{cm^{-2}}$ and $x_{0}=\unit[200]{nm}$.
All energies are measured with respect to $\left(E_{z0}-U_{\mathrm{ee}}\right)_{x\rightarrow-\infty}$.}
\end{figure*}

Initially, we set $U_{x}\left(x\right)=U_{0}\textnormal{sech}^{2}\left(x/x_{0}\right)$,
with $x_{0}$ growing as the length of the QPC increases, while the
QPC barrier height $U_{0}$ is taller when the gate voltage is more
negative. Ballistic transport occurs for $x_{0}$ of the order of
a few hundred nanometers, given that the mean free path is of the
order of several microns\citealp{Wharam1988,Wees1988}. Given that
the confinement length in the $y$-direction ($1/a$) is of the order
of a few tens of nanometers, the adiabatic approximation remains valid
for $x_{0}\gtrsim\unit[100]{nm}$. Meanwhile, the lateral confinement
becomes $U_{\mathrm{well}}\left(x,y\right)=m^{*}\left[\omega_{y}\left(x\right)\right]^{2}y^{2}/2$,
where we assume $\omega_{y}\left(x\right)=\omega_{y,\mathrm{max}}\sqrt{U_{x}\left(x\right)/U_{0}}$.
Once the effective potential $U_{\mathrm{eff}}\left(x,k_{x}\left(x\right)\right)$
(Eq. \eqref{eq:Ueff-x}) is found for sample points along the $x$-axis,
the transmission coefficient $\mathrm{T}\left(E\left(k_{x},\sigma\right)\right)=\mathrm{T}_{\sigma}\left(E\right)$
is calculated by the transfer matrix method\citealp{Jonsson1990}.
Finally, the conductance is obtained using the Landauer formula\citealp{Datta1997},
\begin{equation}
G=\frac{e^{2}}{h}\sum_{\sigma}\int\mathrm{d}E\,\left(-\frac{\partial f_{T}}{\partial E}\right)\mathrm{T}_{\sigma}\left(E\right)\label{eq:Landauer}
\end{equation}

The Fermi level $E_{F}$ is determined by the electron density $n_{2D}$
in the 2DEG far away from the QPC, where the effective potential is
$U_{\mathrm{eff}}\left(x\rightarrow-\infty\right)=\left\langle \hat{T}_{z}\right\rangle +\left\langle U_{A}\right\rangle +U_{\mathrm{ee}}=E_{z0}\left(x\rightarrow-\infty\right)$:
\begin{equation}
E_{F}-U_{\mathrm{eff}\left(x\rightarrow-\infty\right)}=\frac{\pi\hbar^{2}n_{2D}}{m^{*}}\label{eq:EF-vs-n2D}
\end{equation}

Of particular interest is the case of a very low carrier concentration
in the QPC. Then, the effective potential, and therefore the transmission
coefficient, depends on whether or not one of the electrons shares
the same spin with the electron preceding it on the wire. If the electron
spins are different, then the exchange term in the effective potential
drops out and the interaction is anti-ferromagnetic.

\section{Results and discussion}

Figures \ref{fig:6-Ueff-profile-0K}(a)-(d) show the profile of the
effective potential $U_{\mathrm{eff}}\left(E=E_{F}\right)$ at $T=0$
along the $x$-direction when the maximum value of the effective potential,
$U_{\mathrm{eff,max}}$ is, respectively, (a) greater than or (b)
less than $E_{F}$. Here we assume that $n_{2D}=\unit[8.4\times10^{10}]{cm^{-2}}$,
so that for $x\rightarrow-\infty$ the Fermi energy is $\unit[3]{meV}$
above $U_{\mathrm{eff}}$, according to Eq. \eqref{eq:EF-vs-n2D}.
Energies in the plot are measured relative to $\left(E_{z0}-U_{\mathrm{ee}}\right)_{x\rightarrow-\infty}$.
The solid and dashed lines represent the effective potential when
the exchange interaction is either included or ignored. The separation
between the dotted and solid (or dashed) lines corresponds to the
contribution of electron-electron interactions, $U_{\mathrm{ee}}$,
to $U_{\mathrm{eff}}$, which is more significant far away from the
QPC (i.e. where the QPC barrier height, $U_{x}$, drops to zero).

As $U_{x}$ increases near $x=0$, $U_{\mathrm{ee}}$ becomes smaller
because fewer electron states are populated, and therefore $U_{\mathrm{eff}}$
increases less rapidly. In particular, when $U_{\mathrm{eff}}>E_{F}$,
the electron channel is depleted and $U_{\mathrm{ee}}=0$. Figure
\ref{fig:6-Ueff-profile-0K} shows that, at the point where $U_{\mathrm{eff}}=E_{F}$,
there is a kink or shoulder in the effective potential due to the
onset of Coulomb interactions in $U_{\mathrm{eff}}$. Above $E_{F}$,
$U_{\mathrm{eff}}$ varies at the same rate as $U_{x}$. The kink
is not present in Figure \ref{fig:6-Ueff-profile-0K}(b), since the
effective potential in that figure is below $E_{F}$ throughout the
QPC, but $U_{\mathrm{ee}}$ is still smaller near the center of the
QPC when compared to points that are farther away.

Figures \ref{fig:6-Ueff-profile-0K}(c) and (d) show the same plots
for $T=\unit[1]{K}$. The kink in the effective potential is less
pronounced in Fig. \ref{fig:6-Ueff-profile-0K}(c) compared to Fig.
\ref{fig:6-Ueff-profile-0K}(a) because the electron concentration
does not vanish anymore when $U_{\mathrm{eff}}>E_{F}$, and therefore
the Hartree and exchange terms contribute to the effective potential,
especially when $U_{\mathrm{eff}}$ is still very close to $E_{F}$.
Far away from the QPC, the potential profiles are indistinguishable
from their zero-temperature counterparts.

\begin{figure}
\noindent \includegraphics[width=2.8in]{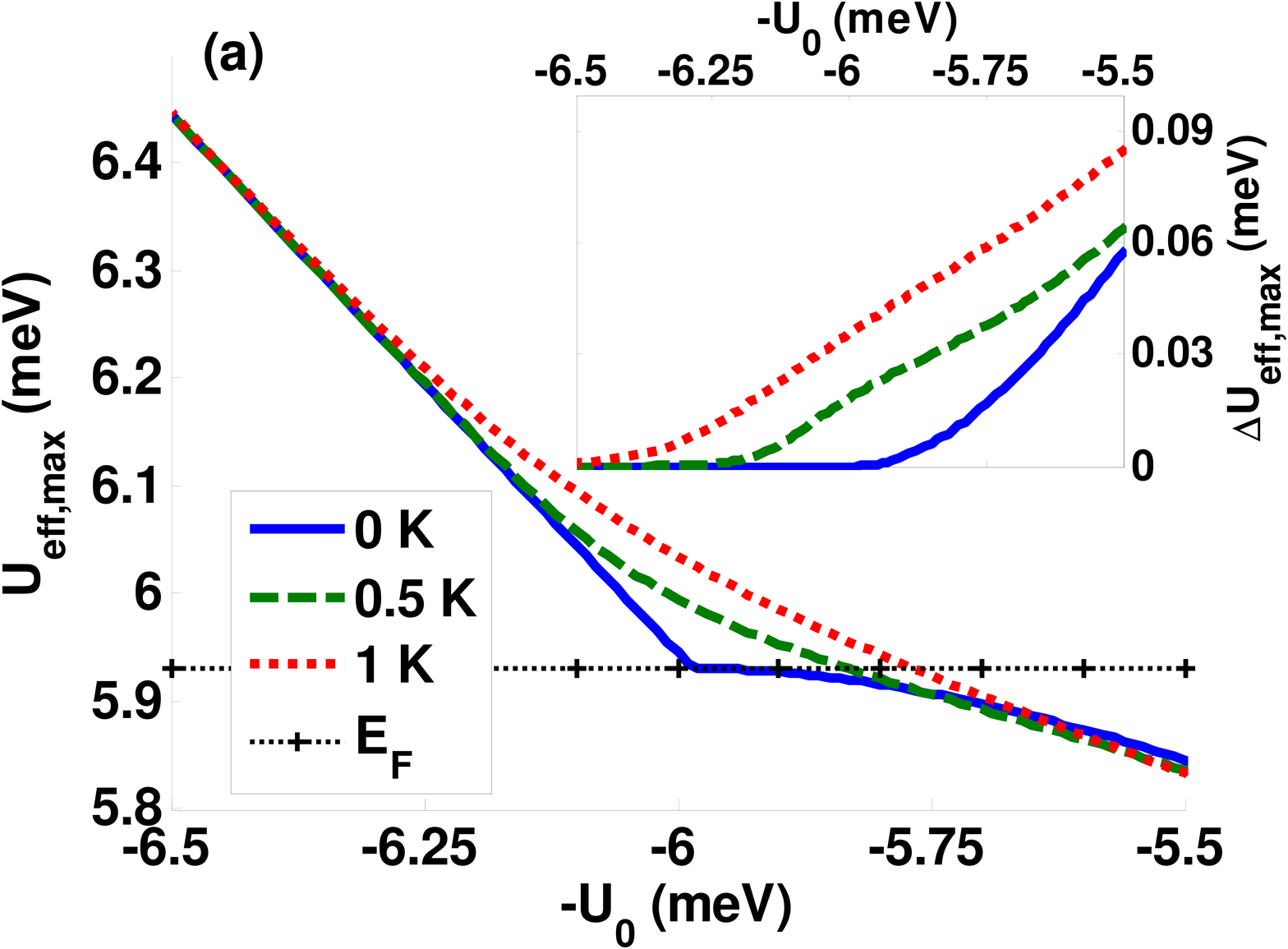}

\noindent \includegraphics[width=2.8in]{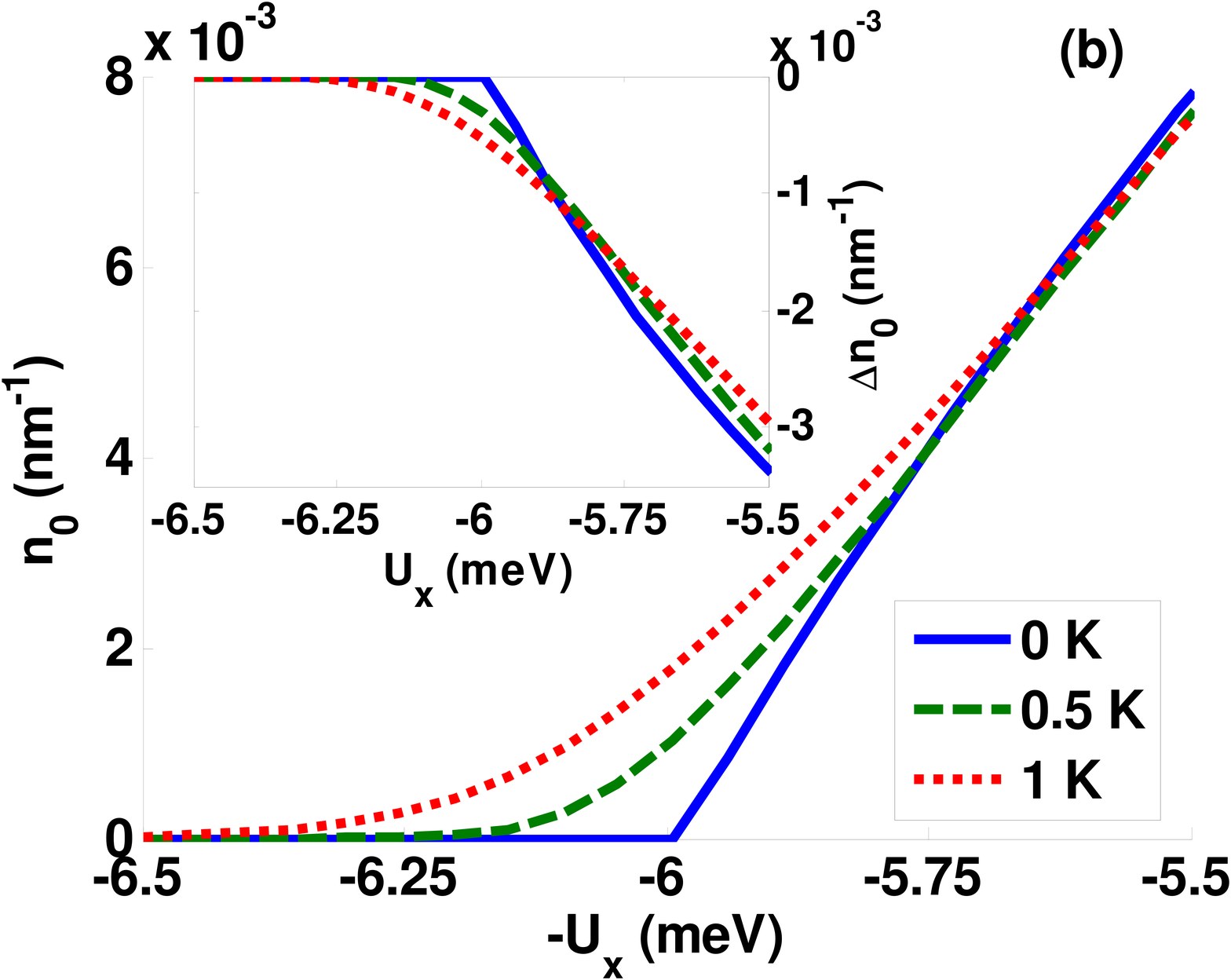}

\caption{\label{fig:7-Ueffmax-and-n0}(Color online) (a) Maximum effective
potential vs. $U_{0}$, including exchange effects, for different
temperatures. Inset: $\Delta U_{\mathrm{eff,max}}=U_{\mathrm{eff,max}}\left(\textnormal{w/o exchange}\right)-U_{\mathrm{eff,max}}\left(\textnormal{w/exchange}\right)$
vs. $U_{0}$. (b) Electron density vs. $U_{x}$, including exchange
effects, for different temperatures. Inset: $\Delta n_{0}=n_{0}\left(\textnormal{w/o exchange}\right)-n_{0}\left(\textnormal{w/exchange}\right)$.
All parameters (except temperature) as in Figure \ref{fig:6-Ueff-profile-0K}.}
\end{figure}

In Figure \ref{fig:7-Ueffmax-and-n0} we plot $U_{\mathrm{eff,max}}$
versus the maximum QPC barrier height ($U_{0}$) for different temperatures.
At zero temperature, the effective potential maximum decreases linearly
with $U_{0}$ above the Fermi energy, i.e. when the QPC is \textquotedblleft{}pinched-off\textquotedblright{}
and there are no electrons in the 1D channel. When $U_{\mathrm{eff,max}}$
crosses the Fermi level at $U_{0}\approx\unit[6]{meV}$, it remains
practically constant (\textquotedblleft{}pinned\textquotedblright{})
over a certain range of $U_{0}$ values while the channel opens. Then,
it decreases at a slow rate for $U_{0}\lesssim\unit[5.75]{meV}$,
when the QPC channel becomes more populated with electrons. This pinning
effect is a consequence of the compressibility peak in the 1D electron
gas, in agreement with previous works\cite{*[{}] [{. We notice, however, that the geometry of the constriction is different from the single QPC considered in our analysis. Additionally, the effective potential in that paper refers to the potential of the detector gate and has a different meaning than in the present work, in which $U_{eff}$ was defined as the 1D potential profile along the constriction.}] Luscher2007,Hirose2001,Ihnatsenka2009}.
It also occurs when $U_{\mathrm{eff}}$ crosses the Fermi level on
both sides of the QPC when the 1D channel is still closed, and is
responsible for the \textquotedblleft{}kink\textquotedblright{} observed
in the barrier profile (Fig. \ref{fig:6-Ueff-profile-0K}(a)). The
pinning is less effective at non-zero temperatures because the electron
density in the QPC is not zero. The dependence of $n_{0}$ on $U_{x}$
is shown in Figure \ref{fig:7-Ueffmax-and-n0}(b). As predicted by
Eq. \eqref{eq:n0-0K}, decreasing $U_{x}$ past the critical point
at $U_{x}\approx\unit[6]{meV}$ leads to an increase in $n_{0}$.
Meanwhile, $n_{0}$ increases with rising temperature near the effective
potential crossover, as expected.

The changes in $U_{\mathrm{eff,max}}$ and $n_{0}$ when exchange
effects are neglected are shown in the insets of Figures \ref{fig:7-Ueffmax-and-n0}(a)
and (b). As can be seen in Fig. \ref{fig:7-Ueffmax-and-n0}(a), neglecting
the (negative) exchange term leads to an increase in the maximum effective
potential below $E_{F}$, as explained in the comments of Fig. \ref{fig:6-Ueff-profile-0K}.
In addition, $U_{\mathrm{eff,max}}$ also becomes less sensitive to
variations in $U_{0}$, which indicates that the pinning of the effective
potential is enhanced by the absence of the exchange interaction.
The increase in $U_{\mathrm{eff,max}}$ becomes even more pronounced
when temperature is increased. Since the effective potential is higher,
the carrier concentration in the absence of exchange effects is lower
relative to the case with exchange, as shown in the inset of Fig.
\ref{fig:7-Ueffmax-and-n0}(b).

\begin{figure}
\noindent \includegraphics[width=2.8in]{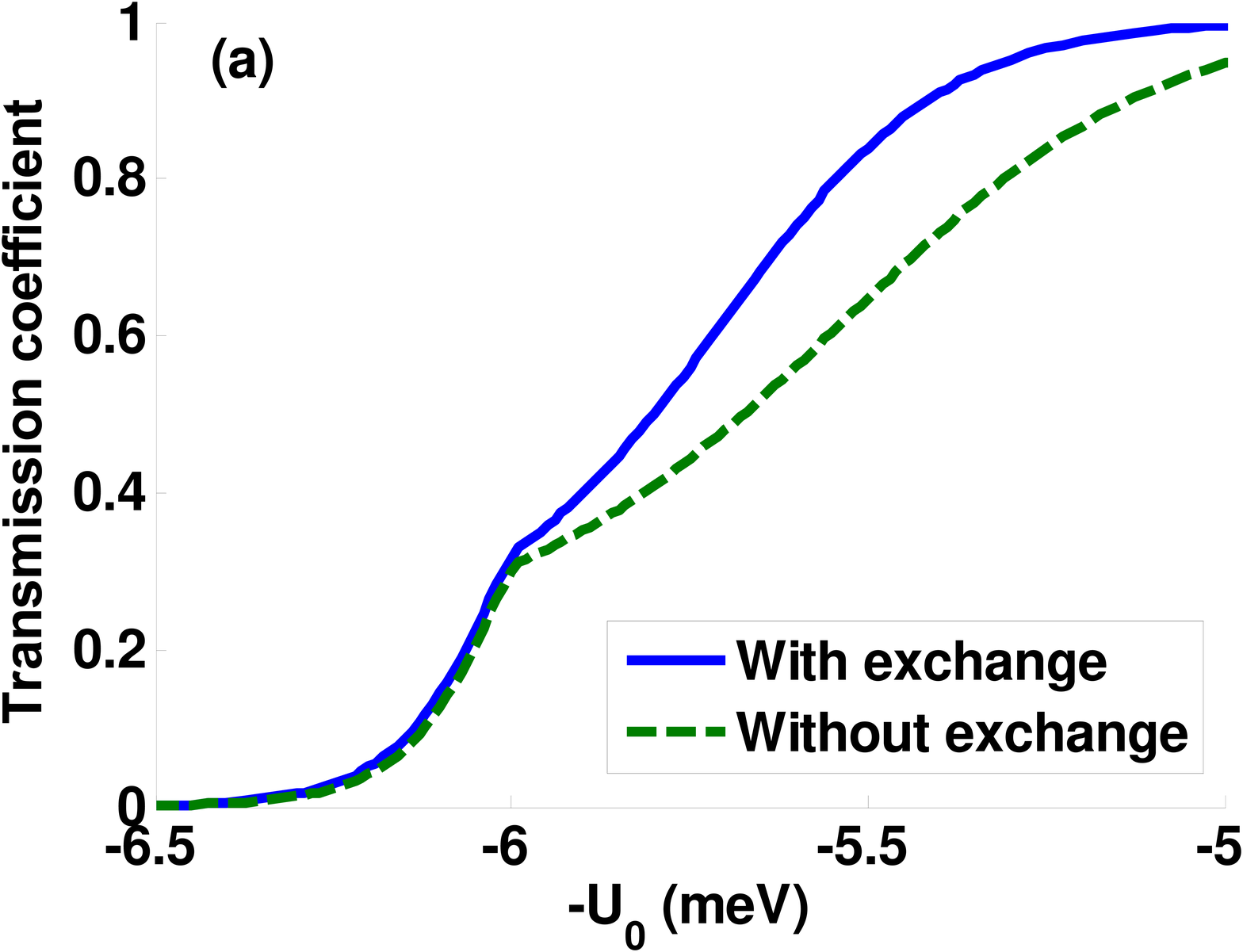}

\noindent \includegraphics[width=2.8in]{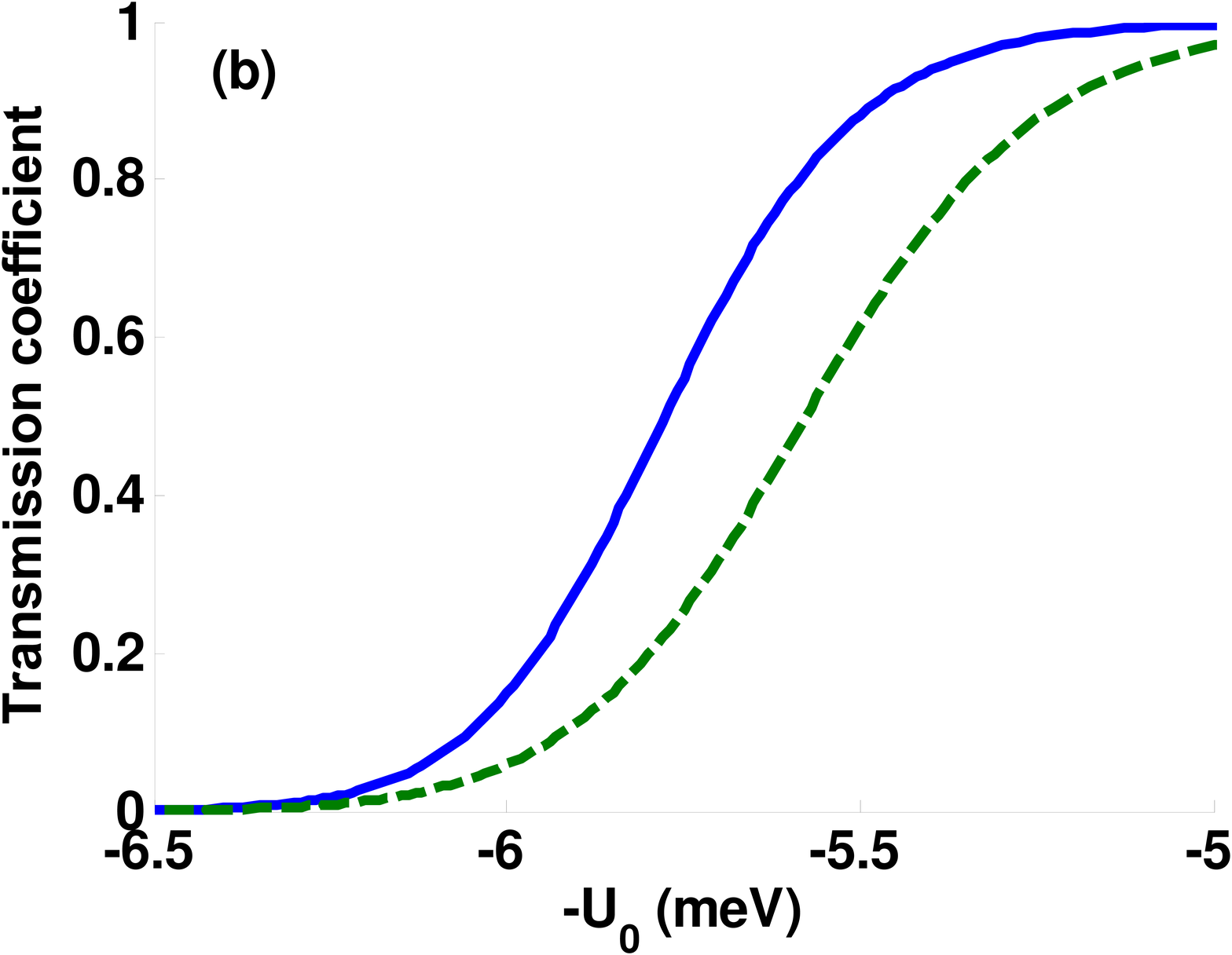}

\caption{\label{fig:8-trcoeff}(Color online) Transmission coefficient vs.
$U_{0}$ for (a) $T=\unit[0]{K}$; (b) $T=\unit[1]{K}$, showing the
curves in the presence and in the absence of the exchange interaction.
All other parameters as in Figure \ref{fig:6-Ueff-profile-0K}.}
\end{figure}

A plot of the transmission coefficient as a function of $U_{0}$ at
$T=0$ (Figure \ref{fig:8-trcoeff}(a)) reveals a kink or anomaly
close to 0.35, when $U_{0}$ reaches the critical point of $\sim\unit[6]{meV}$
and $U_{\mathrm{eff,max}}$ crosses $E_{F}$. This kink is a consequence
of the pinning of the effective potential just below $E_{F}$ and
is more pronounced when the exchange interaction is neglected (dashed
curve) since the effective potential is enhanced, as mentioned in
our discussion of the insets in Figs. \ref{fig:6-Ueff-profile-0K}
and \ref{fig:7-Ueffmax-and-n0}. In Figure \ref{fig:8-trcoeff}(b),
we display the transmission coefficient at $T=\unit[1]{K}$ and show
that the anomaly disappears with increasing temperature. This is caused
by the thermal smearing of the effective potential pinning, as shown
in Fig. \ref{fig:7-Ueffmax-and-n0}. At zero temperature, when the
gate voltage is sufficiently negative, the transmission coefficient
drops rapidly because of the sudden increase in the height $U_{\mathrm{eff,max}}$
of the effective potential (Figures \ref{fig:7-Ueffmax-and-n0}(a)
and (c)). However, due to the modulation of the potential barrier
with increasing temperature, the changes in $U_{\mathrm{eff,max}}$
are smoother and the transmission coefficient does not change as abruptly,
thus softening the anomaly. By neglecting the (negative) exchange
interaction (dashed curve), the effective potential barrier is taller,
thereby resulting in a lower transmission coefficient, as can be seen
both in Figure \ref{fig:8-trcoeff}(a) (for $U_{0}$ below the critical
value of $\unit[6]{meV}$) and \ref{fig:8-trcoeff}(b).

\begin{figure}
\noindent \includegraphics[width=2.8in]{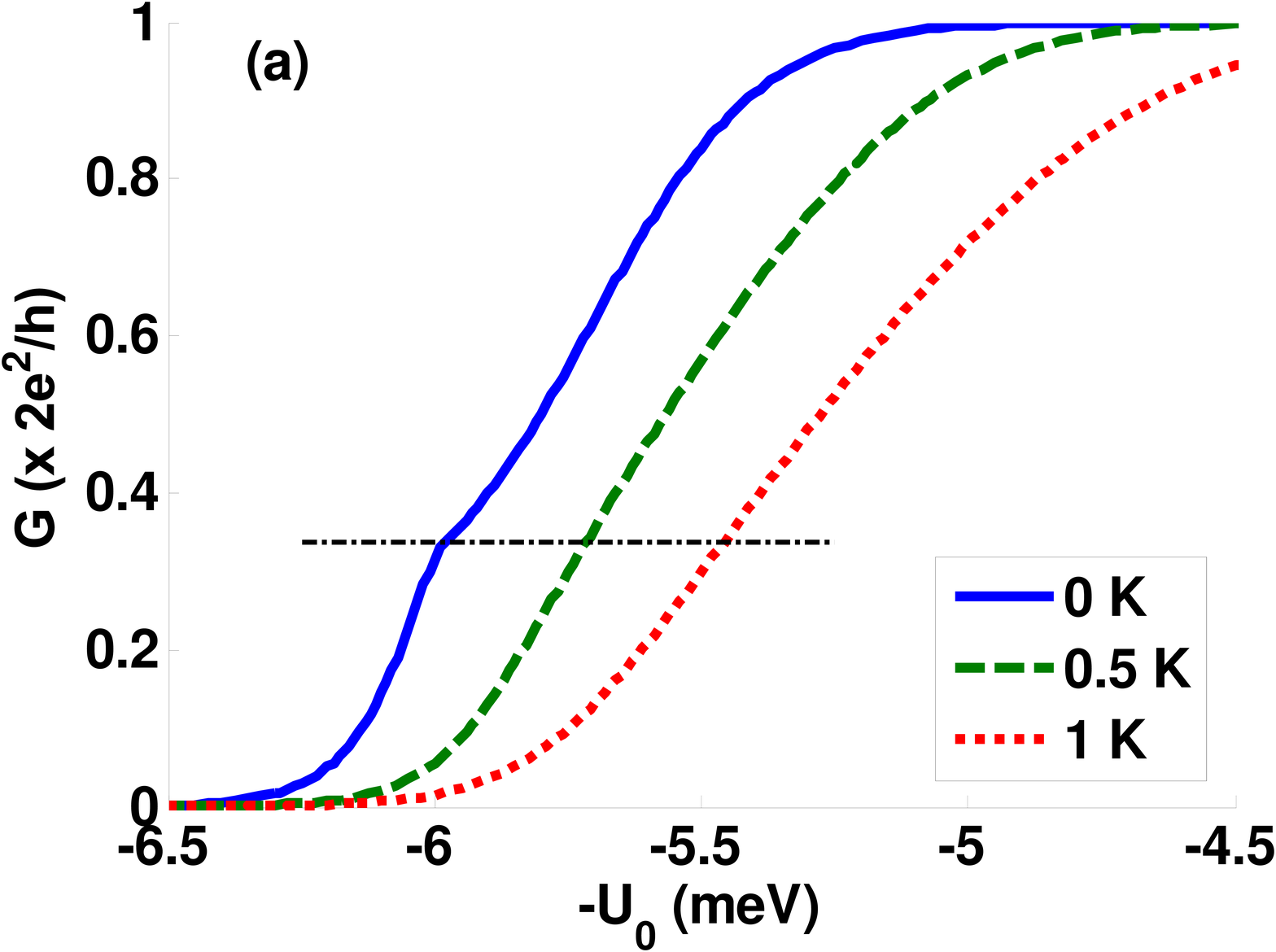}

\noindent \includegraphics[width=2.8in]{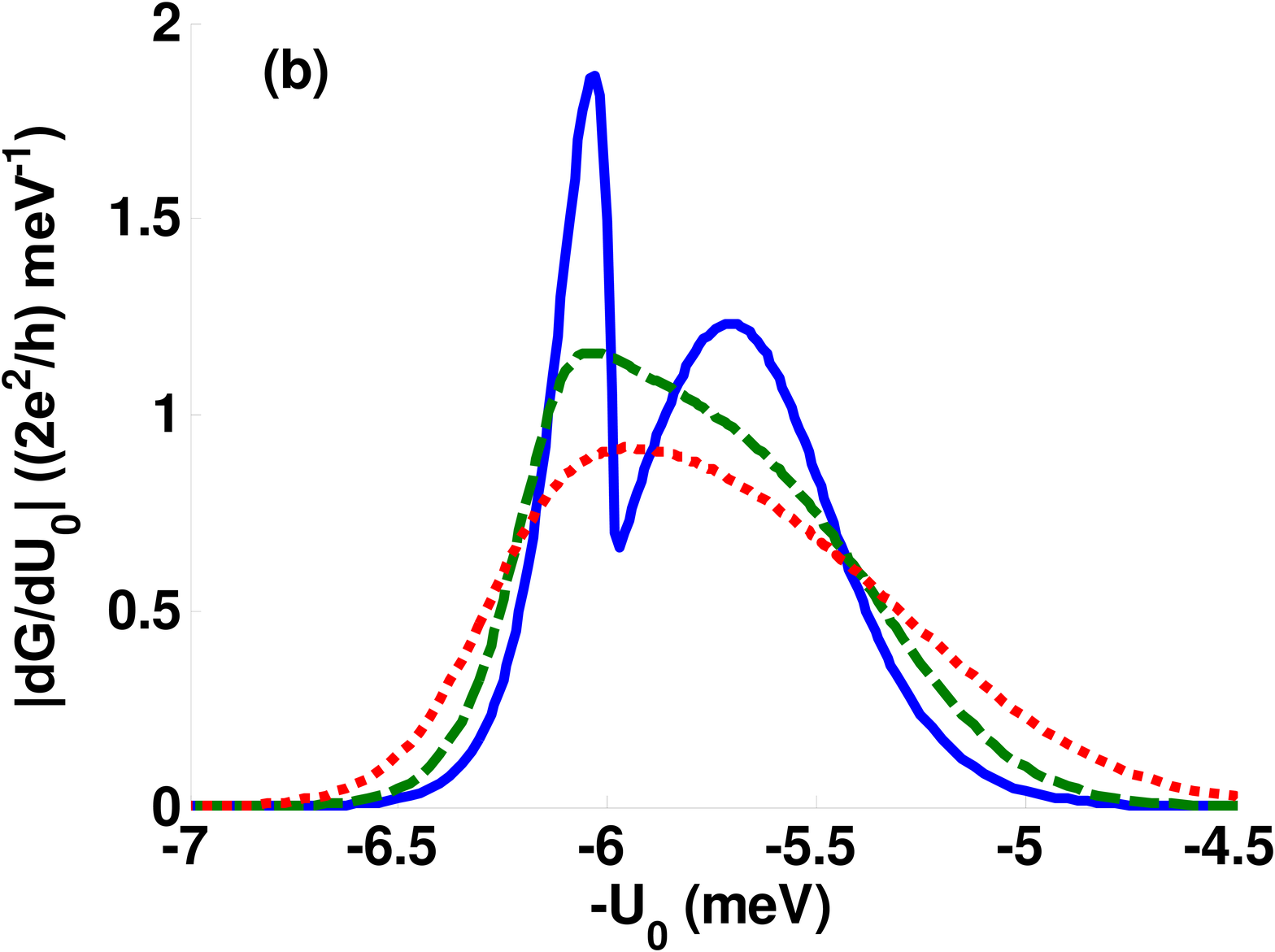}

\caption{\label{fig:9-conductance}(Color online) (a) Conductance vs. $U_{0}$
for different temperatures. For clarity, each successive curve is
shifted to the right by $\unit[0.25]{meV}$. The dot-dashed line indicates
the location of the kink. (b) Slope of the conductance, $\left|dG/dU_{0}\right|$,
vs. $U_{0}$. Exchange effects are included. All parameters as in
Figure \ref{fig:6-Ueff-profile-0K}.}
\end{figure}

Figure \ref{fig:9-conductance}(a) shows the QPC conductance as a
function of $U_{0}$ for different temperatures. Here we included
exchange effects on the barrier. At $T=\unit[0]{K}$, the kink in
the transmission coefficient at $G\sim0.35G_{0}$ is well reproduced
in the conductance, but is smeared out as temperature increases, even
by a few tenths of a kelvin. This thermal smearing is confirmed in
Figure \ref{fig:9-conductance}(b), where we plot the slope of the
conductance as a function of $U_{0}$. At $T=\unit[0]{K}$, the conductance
slope exhibits a double peak with a sharp maximum before $U_{0}=\unit[6]{meV}$,
followed by a broader and lower maximum, which indicates that the
kink in the conductance is not simply a slope change, but is rather
due to the onset of the 1D compressibility of the electron gas. However,
as temperature increases the dip in the double peak structure disappears
to leave a single and broad peak.

\begin{figure}
\noindent \includegraphics[width=2.8in]{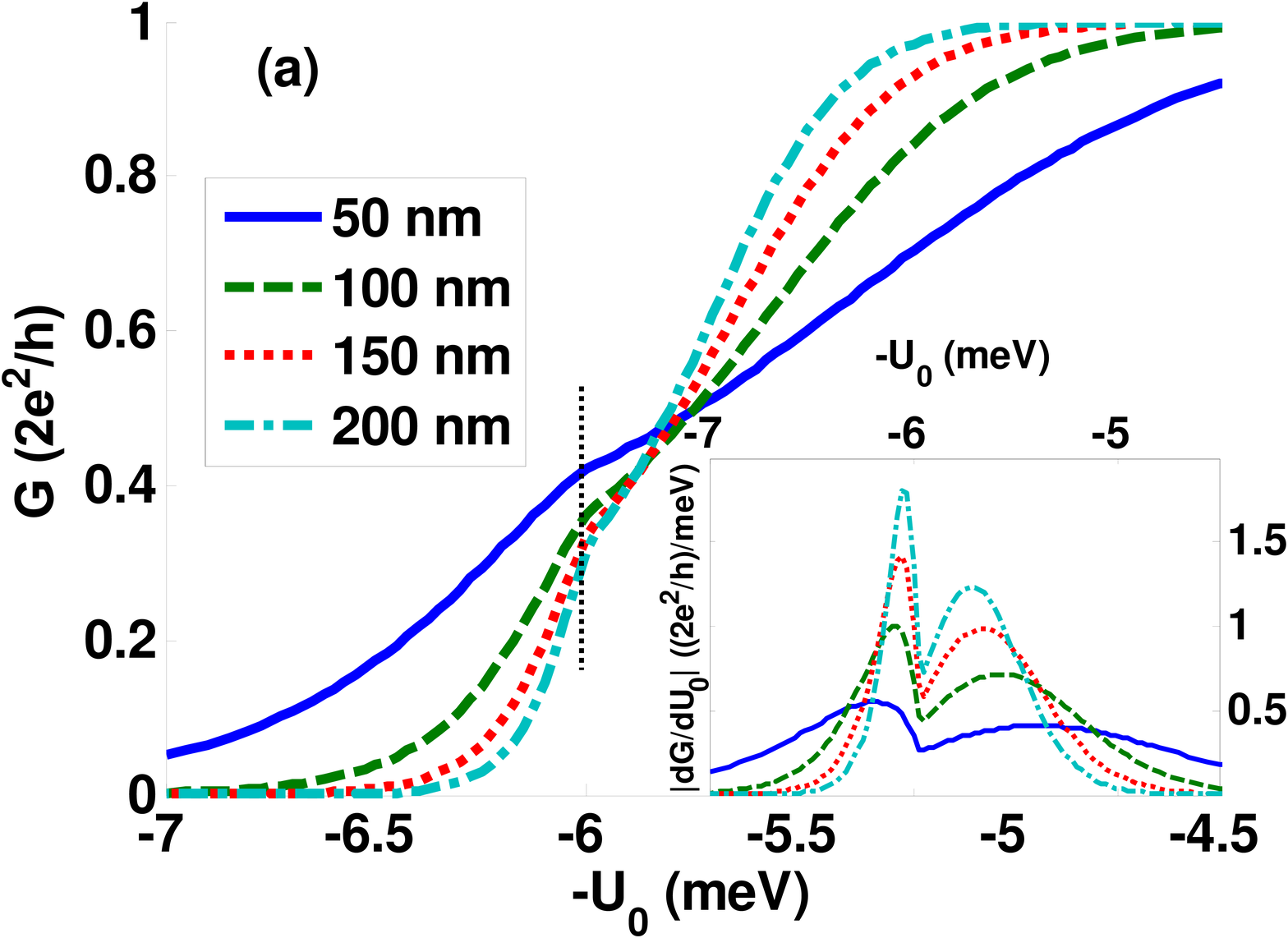}

\noindent \includegraphics[width=2.8in]{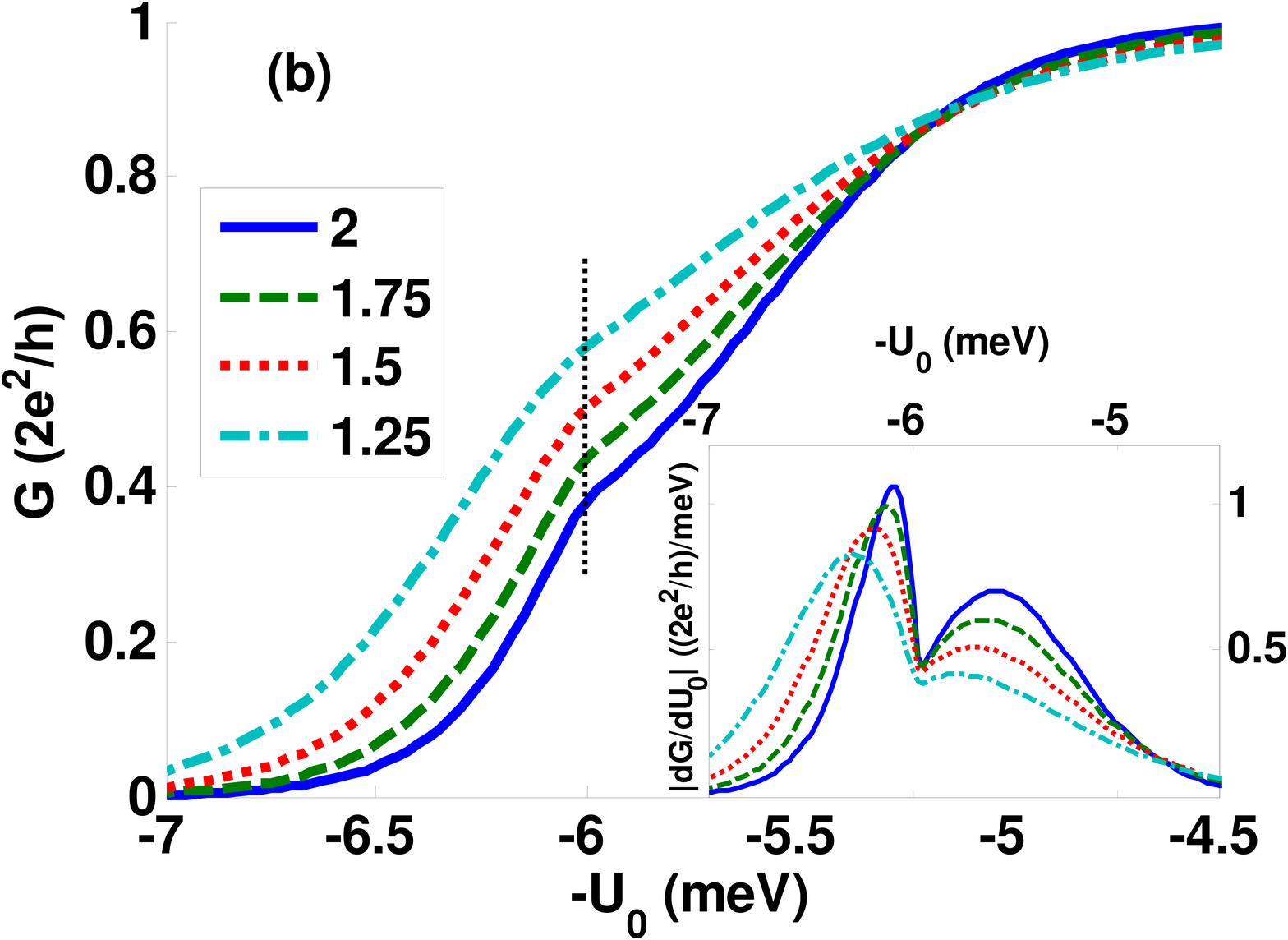}

\caption{\label{fig:10-Cond-for-diff-potentials}(Color online) (a) Conductance
at $T=0$ vs. $U_{0}$ for: (a) $U_{x}=U_{0}\textnormal{sech}^{2}\left(x/x_{0}\right)$,
for different $x_{0}$; (b) $U_{x}=U_{0}\left[1+\left(x/x_{0}\right)^{N}\right]^{-1}$,
for different $N$. Exchange effects are included. All parameters
(apart from $x_{0}$ in (a)) are the same as in Figure \ref{fig:6-Ueff-profile-0K}.
The thin vertical dotted line shows the location of the kink. Insets:
slope of the conductance vs. $U_{0}$.}
\end{figure}

In Figure \ref{fig:10-Cond-for-diff-potentials}(a), we display the
sensitivity of the $T=\unit[0]{K}$ conductance as a function of the
QPC length, which consists in varying $x_{0}$, the characteristic
length of the potential barrier (defined as $U_{x}=U_{0}\textnormal{sech}^{2}\left(x/x_{0}\right)$).
As can be seen, a decrease in $x_{0}$, corresponding to a shorter
QPC, leads to a spread of the conductance with a lower slope over
a broader $U_{0}$ potential range, and shifts the kink anomaly upwards,
towards $G\sim0.4G_{0}$. At the same time, the kink becomes softer
and broader (Fig. \ref{fig:10-Cond-for-diff-potentials}(a), inset).
This upward displacement of the kink is caused by the narrower barrier
for which the transmission is enhanced at low energy compared to the
conductance in longer QPCs. A more dramatic effect on the conductance
is observed if the $U_{x}$-potential profile changes. In Figure \ref{fig:10-Cond-for-diff-potentials}(b),
we use a barrier potential of the form $U_{x}=U_{0}\left[1+\left(x/x_{0}\right)^{N}\right]^{-1}$
(where $N$ is a positive exponent) to calculate the QPC conductance
at $T=\unit[0]{K}$. Compared to the $\textnormal{sech}^{2}\left(x\right)$
potential shape, this function has a sharper drop at $x=0$ and a
longer tail for $x>x_{0}$, especially when $N$ is small. As can
be seen, a decrease in the value of the exponent $N$ leads to an
upwards displacement of the anomaly since for $N=2$ the anomaly occurs
at $G\sim0.4G_{0}$, whereas for $N=1.25$ it rises up to $G\sim0.6G_{0}$.
The transmission is higher in this case compared to higher-$N$ potentials
over the whole $U_{0}$ energy range. At the same time, the anomaly
broadens and shifts to lower $U_{0}$ potential values (Fig. \ref{fig:10-Cond-for-diff-potentials}(b),
inset).

\section{Conclusions}

We developed a 3D quantum mechanical model for near-equilibrium ballistic
transport through a constriction in a 2D GaAs/AlGaAs electron gas
by using a self-consistent variational approach. We were able to define
an effective 1D potential ($U_{\mathrm{eff}}$) for the constriction,
which takes into account the static potential arising from fixed acceptor
charges, many-body effects, and the confinement and barrier potentials.
Away from the constriction, the model reduces to that of a 2DEG with
a fixed Fermi level. In the constriction, the gate-induced potential
leads to a downward shift of the 2DEG subbands. Our model, based on
a transmission matrix technique, predicts an anomaly in the range
$0.3G_{0}<G<0.6G_{0}$, during the rise to the first conductance plateau
at $T=\unit[0]{K}$. This anomaly is caused by a change of slope in
the variation of the effective potential with the external gate voltage
due to the charging of the 1D channel at the compressibility peak,
in agreement with previous observations\citealp{Luscher2007,Hirose2001,Ihnatsenka2009}.
Our model also predicts an anomaly enhancement in the case of anti-ferromagnetic
interaction in the QPC as the attractive exchange interaction lowers
the potential barrier, thereby disqualifying the presence of any spontaneous
spin polarization. Our main result, however, is that the exact conductance
value for the anomaly depends on the length of the QPC and the shape
of the potential barrier, i.e. long QPCs make the conductance sharper,
while sharper potentials lead to a higher value of the conductance
kink. These findings tend to be in agreement with experimental observations
that show a shift of the 0.7 plateau toward higher conductance values
under higher source\textendash{}drain biases\citealp{Cronenwett2002},
which also sharpens the QPC potential. However, our self-consistent
model also shows the barrier sensitivity to temperature that smears
out the anomaly, unlike the enhancement of the 0.7 feature with increasing
temperature that is observed experimentally\citealp{Thomas1996,Nuttinck2000,Cronenwett2002}.
We believe this result is due to the approximation of a smooth QPC
barrier, which may not well reproduce the potential landscape characterized
by the AlGaAs substitutional disorder at the GaAs/AlGaAs interface\citealp{Thean1997},
nor any possible bound states required for the occurrence of a Kondo-like
effect\citealp{Cronenwett2002,Rejec2006}. These issues, as well as
the investigation of far-from-equilibrium transport in QPCs, will
be the subject of a forthcoming paper.
\begin{acknowledgments}
A.X. Sánchez thanks the Department of Physics at the University of
Illinois at Urbana-Champaign for their continued support during his
studies.
\end{acknowledgments}
\bibliographystyle{apsrev4-1}
\bibliography{C:/Documents/UIUC/Research/References/AllReferences}

\end{document}